\documentclass[preprint,3p,times]{elsarticle}

\usepackage{amssymb}
\usepackage{comment}
\usepackage[figuresright]{rotating}
\usepackage{subcaption}
\usepackage{graphicx}
\usepackage{multirow}
\usepackage{colortbl}
\usepackage{amsmath}
\usepackage{hyperref}
\usepackage{bigstrut}
\begin{document}

\begin{frontmatter}



%% e.g. \dochead{17th International Conference on Dynamical Processes in Excited States of Solids}

\title{LPASS: Linear Probes as Stepping Stones for vulnerability detection using compressed LLMs}


\address[label1]{Universidad Carlos III de Madrid, Leganes, Madrid}
\address[label2]{Inria de Saclay, Palaiseau, France}

\author[label1]{Luis Ibanez-Lissen}
\address[label3]{Institut Politechnique de Paris, Palaiseau, France}
\address[label4]{Corresponding author}

\author[label1,label3,label4]{Lorena Gonzalez-Manzano}

\author[label1,label2]{Jose Maria de Fuentes}

\author[label2]{Nicolas Anciaux}

\begin{abstract}
Large Language Models (LLMs) are being extensively used for cybersecurity purposes. One of them is the detection of vulnerable codes. For the sake of efficiency and effectiveness,  compression and fine-tuning techniques are being developed, respectively. However, they involve spending substantial computational efforts. In this vein, we analyse how Linear Probes (LPs) can be used to provide an estimation on the performance of a compressed LLM at an early phase -- before fine-tuning. We also show their suitability to set the cut-off point when applying layer pruning compression. Our approach, dubbed $LPASS$, is applied in BERT and Gemma for the detection of 12 of MITRE's Top 25 most dangerous vulnerabilities on 480k C/C++ samples. LPs can be computed in 142.97 s. and provide key findings: (1) 33.3 \% and 72.2\% of layers can be removed, respectively, with no precision loss; (2) they provide an early estimate of the post-fine-tuning and post-compression model effectiveness, with 3\% and 8.68\% as the lowest and average precision errors, respectively. $LPASS$-based LLMs outperform the state of the art, reaching 86.9\% of accuracy in multi-class vulnerability detection. Interestingly, $LPASS$-based compressed versions of Gemma outperform the original ones by 1.6\% of F1-score at a maximum while saving 29.4 \% and 23.8\% of training and inference time and 42.98\% of model size.
\end{abstract}


\end{frontmatter}

% \linenumbers

\section{Introduction}
The wide variety of devices, systems and networks favors the emergence of vulnerabilities, defined as weaknesses in a system or security control\footnote{https://csrc.nist.gov/glossary/term/vulnerability , last access September 2024}. Lots of research works have been carried out to counter them. Particularly, multiple works have focused on vulnerability detection \cite{chen2023diversevul,du2024vulnerability,ding2024vulnerability}, that is, whether a piece of software is vulnerable or not, whereas others aim to identify them, that is, which vulnerability is present 
\cite{fu2022linevul,steenhoek2023empirical,hanif2022vulberta}. This latter problem is more challenging as it requires a fine-grained knowledge of all vulnerabilities to tell them apart \cite{zhout2023devil}. Therefore, in this paper we concentrate on this issue. 

\textbf{Context.} The use of Artificial Intelligence (AI) has been a constant in this field. Many works apply traditional AI algorithms such as support vector machines \cite{lomio2022just,chernis2018machine}, K-nearest neighbor \cite{lomio2022just,chernis2018machine} or graph neural networks \cite{tang2023csgvd,hin2022linevd}, once having extracted code features like the program dependency graph \cite{tang2023csgvd,hin2022linevd}, the control flow graph \cite{do2023novel} or the number of lines of code \cite{lomio2022just}, among others. 

However, since the {\color{black} appearance} of Large Language Models (LLMs), which are based on using deep learning techniques to manage massive amounts of data, novel vulnerability detection systems {\color{black} have appeared \cite{sheng2025large}}. The main advantage of these models is their ability to manage complex tasks, but the required computational cost becomes a burden. Some of these models apply millions or even billions of parameters whose tuning and execution involves a high cost in terms of time and resources.

New research lines try to reduce LLMs, speeding up the classification process and reducing the amount of used computing resources. Techniques such as pruning focus on reducing the amount of model parameters \cite{gholami2022survey} or network components \cite{he2023structured}. On the other hand, quantization is another well-known approach that reduces the size of each parameter. However, compression methods incur in accuracy losses which have to be minimized \cite{zhou2024survey}. In the context of vulnerability detection, to the best of authors' knowledge only \cite{shi2022compressing}, \cite{shi2024greening} have applied model compression. They apply knowledge distillation \cite{hinton2015distilling} to compress CodeBERT and GraphCodeBERT, reaching promising inference speeds and efficiency. However, they focus on a binary classification problem (i.e., vulnerable / non-vulnerable) with an accuracy of 59.9 \% -- only 10\% over a random guess.

% accuracy  {\color{red} Luis: read https://openreview.net/pdf?id=nMbWsXPUVL  Moreover, add any ref to support (1)the effect of model compression in other fields, (2) why these results are not enough for our field (which are the specificities of vul det that justify our study)}. Despite existing techniques for LLM reduction, they have not been applied in the vulnerability detection field. {\color{red} Luis: leer estos dos papers \cite{shi2022compressing}\cite{10.1145/3639475.3640097}, que SI han usado compresión para detección de vulnerabilidades y de hecho ya dicen cosas muy fuertes como que se puede reducir 160 veces el tamaño del modelo con una perdida del 1 por ciento de accuracy. Justificar por que otro paper. Anadir en tabla de RW.}

\textbf{Motivation.}  Efficiency in vulnerability detection is of outmost relevance considering the increasing pace of software generation. For the sake of illustration, Google Play counts on 2.61 billion apps nowadays \footnote{https://www.businessofapps.com/data/google-play-statistics/, last access 20 june 2024.}. At the same time, not only the amount of vulnerabilities, but also their severity, have been steadily growing in the last decade. According to CVEdetails\footnote{https://www.cvedetails.com/, last access September 2024}, the amount of vulnerabilities with a severity ranked between 7 and 10 raised from 1.9k in 2014 to 16.7k in 2023. Therefore, a vulnerability detection mechanism streamlined with the publication process would be desirable. Interestingly, identifying which is the vulnerability at stake enables providing a suitable response -- while dangerous vulnerabilities may require stronger controls, irrelevant ones may simply raise a warning before distributing the analysed pieces of software.

As noted in \cite{wang2023energy}, fine-tuning of models involves substantial energy expenses. For the sake of illustration, pre-training and fine-tuning Meta’s LLaMA model could cause up to 2.76 MtCO$_2$-eq emissions, equivalent to the total pollution caused by manufacturing one dose of COVID-19 vaccine for all humans on Earth \cite{jiang2024preventing}. Therefore, saving resources related to the model fine-tuning and compression is paramount.

The closest effort to ours, namely Chen et al. \cite{chen2018shallowing} is focused on compressing visual models, thus unrelated to vulnerability detection. They proposed using feature representations in convolutional neural networks to identify layers with high weight overlap \textit{after training}, aiming to reduce the model size.  In contrast, our approach seeks to compress pretrained LLMs by determining which layers of a model provide valuable information for a specific task \textit{before any fine-tuning or further training}. Therefore, although Chen et al.’s method can save resources post-training, it requires training the entire model beforehand. Our approach, on the other hand, reduces the model size prior to fine-tuning, minimizing the computational resources needed for that task. 

%\textbf{Contribution.} This paper presents an approach (dubbed $LPASS$) to generate light LLMs for vulnerability detection. It leverages linear classifier probes \cite{probes} to get insights of the internal model status. Interestingly, no previous work (in any domain) has used this type of information to guide the model compression. It must be noted that linear classifier probes offer a first layer of explainability of the model by providing interpretability. As such, $LPASS$ can be regarded as a first step for a explainable compression method, which has been recently pointed out as an open issue \cite{zhu2023survey}. $LPASS$ is tested using 3 widespread datasets of C/C++ and a couple of the most used LLMs.  number\footnote{\url{https://cwe.mitre.org/index.html}, last access June 2024.}, dealing with a multi-classification problem.

\textbf{Research question and contribution.} %This paper explores the use of model compression techniques in the field of vulnerability detection. We characterize their impact in the model performance. However, 
Given the computational cost of model fine-tuning and compression, our research question is: \textit{Can we predict their impact to make informed decisions on their use and save resources}? %In this vein,
To address this, 
%we aim to get some early estimates on the model performance. To this end, 
we %resort to 
leverage linear classifier probes (LPs) \cite{alain2016understanding} to gain %get
early insights of the internal model status that can be valuable to guide the use of fine-tuning and compression techniques and estimate their effect in performance. Our approach, dubbed $LPASS$ (LPs As Stepping Stones) is inspired by recent efforts that apply LPs to analyse the knowledge captured at varying depths within LLMs \cite{jin2024exploring}, as well as the internal uncertainty \cite{ahdritz2024distinguishing}. LPs are a first layer of explainability by providing interpretability. As such, our use of LPs can be regarded as a first step for a explainable compression method, which has been recently pointed out as an open issue \cite{zhu2023survey}. 

%To further clarify our contribution, it is interesting to state what it is \textit{not}. The focus is not on creating a new compression technique, as there are myriads of them \cite{zhu2023survey}. We are also not the first ones in showing how compressed LLMs may be applied for vulnerability detection, as Shi et al. did \cite{shi2022compressing}. Thus, our aim is to help on using compression techniques in an informed manner. 

The list of contributions is as follows:
\begin{itemize}
    %\item We propose $LPASS$ an approach for model compression considering model-internal data extracted by linear classifier probes.
    %\item We analyse the impact of model compression techniques, namely layer pruning and quantization, in the field of multiclass vulnerability detection.   

    \item  We adopt linear probes (LPs) in vulnerability detection for 1) determining the cut-off point when applying layer pruning and 2) estimating the effectiveness and performance of fine-tuned and compressed models. %We adopt linear classifier probes (LCPs) \cite{?} to guide the use of LLM compression techniques in the field of vulnerability detection. LCPs are applied on two well-known compression techniques, namely structural pruning \cite{?} and quantization \cite{?}. This leads to a compression procedure with some degree of interpretability that has never been explored so far.
    \item We test $LPASS$ in two well-known LLMs, namely Bert \cite{devlin2018bert} and Gemma \cite{team2024gemma}, compressed by means of layer pruning and quantization \cite{jacob2018quantization}. Bert is selected for being a common choice in previous works, whereas Gemma is a recent, state-of-the-art alternative. Three representative datasets are applied, namely  DiverseVul \cite{chen2023diversevul}, Big-Vul \cite{fan2020ac} and PrimeVul \cite{ding2024vulnerability}. We focus on 12 of the most dangerous vulnerabilities according to MITRE Top 25 \cite{christey2013common}. Our compressed models outperform the state-of-the-art while exhibiting promising performance features.
    \item We release our experimental materials to foster further research.
\end{itemize}

The remaining of this paper is as follows. Section \ref{sec:Preliminaries} gives the background. Section \ref{sec:foundations} describes the foundations of $LPASS$, whereas Section \ref{sec:core} details the approach. Section \ref{sec:assess} describes its assessment. Section \ref{RW} shows the related work. Lastly, Section \ref{sec:conclusion} concludes the paper and points out future work directions.

\section{Preliminaries}
\label{sec:Preliminaries}
In this section, model compression techniques, the notion of linear probes and the Common Weakness Enumeration (CWE) approach for naming vulnerabilities are introduced. 
\subsection{Model compression}
\label{sec:modelComp}

There are several techniques available to reduce the size of models and lower computational resource requirements. Interested readers may refer to \cite{zhu2023survey}. The following three are predominant.

One such technique is knowledge distillation \cite{hinton2015distilling}, where the responses of a larger model (the teacher) are used to guide the training of a smaller version of the original model (the student).

{
\color{black}
Another commonly used technique is pruning, which was previously referred to as 'neural network pruning' \cite{lecun1989optimal,han2015learning},} which aims at reducing model size. Pruning can be done either by modifying the original model architecture (structured pruning) or by removing individual weights and activations (unstructured pruning). This technique reduces the model's complexity and the memory needed to run and store the models.

Quantization, on the other hand, focuses on reducing the model size and computational requirements by mapping continuous infinite numbers to a smaller set of discrete finite numbers. It involves converting weights stored in high-precision values to lower-precision data types. For example, 32 bits weights can be mapped to 8-bit integers in the range [-128, 127] or to 4-bit integers in the range [-8, 7]. It minimizes the number of required bits and the precision of the computations while trying to maximize accuracy, either post-training or after training \cite{gholami2022survey}.  According to Marchisio et al. \cite{marchisio2024does} the effect of quantization is dependent on the language or context. Thus, anticipating its effects is far from straightforward.

%{
%\color{black}Quantization is the process of mapping continuous infinite numbers to a smaller set of discrete finite numbers. This technique minimizes the number of required bits while maximizing the efficiency of computations, making it a common approach in the design of efficient neural network models \cite{gholami2022survey}. }

\subsection{Linear probes}
\label{LP}

Linear Classifier Probes, hereinafter Linear Probes (LP), are simple classifiers that contribute to deep learning models explainability efforts by providing insights into how the model processes information internally \cite{alain2016understanding}. These LPs are used to make predictions over the hidden states of the models, trying to predict or identify if some specific information is correctly represented within them.
For LLMs, a LP classifier is typically placed after each layer of the network and takes the hidden states as input $X$ and predicts a simple characteristic $Y$ (e.g. predict the number of lines of a piece of code). They are trained on probing datasets designed to predict expected characteristics that are predefined and known in advance, giving a sense of how different layers of the model encode and retain the expected information. {\color{black}They have previously been employed to enhance explainability in tasks such as document ranking \cite{wallat2023probing}, and to shed light on model behavior, including hallucinations and the internal representation of code and general knowledge \cite{duan2024llms,karmakar2021pre,steenhoek2023language,jin2024exploring,manigrasso2024probing}.
 Beyond their use for understanding representations, linear probes have also been applied to directly improve model performance—by identifying task-relevant components within the model \cite{hoscilowicz2024non,abbas2024enhancing}—and to mitigate undesirable behaviors such as sycophancy \cite{papadatos2024linear}.

}

%Other studies use LPs to demonstrate that LLMs contain internal representations of different uncertainty in the absence of ground truth \cite{ahdritz2024distinguishing}.

\subsection{Common Weakness Enumeration (CWE)}
\label{CWE}

The Common Weakness Enumeration (CWE) \cite{christey2013common} is a community-driven classification and categorization system designed to identify potential common software and hardware vulnerabilities. It defines families of taxonomies that encompass vulnerabilities, which are interrelated or serve as the basis for higher-level abstract classes, by assigning a unique identifier and a potential damage score. This information is subsequently used to rank common flaws and errors made by developers.

For instance, the MITRE Top 25\footnote{https://cwe.mitre.org/top25/} is a list updated three times per year that highlights the most common and impactful software weaknesses, which are often easy for attackers to exploit.

\section{$LPASS$ foundations}
\label{sec:foundations}
Section \ref{sec:overview} provides an overview of the proposal to then introduce the pursued goals in Section \ref{sec:goals}.

\begin{figure}[th]
    \centering
    \includegraphics[width=0.8\textwidth]{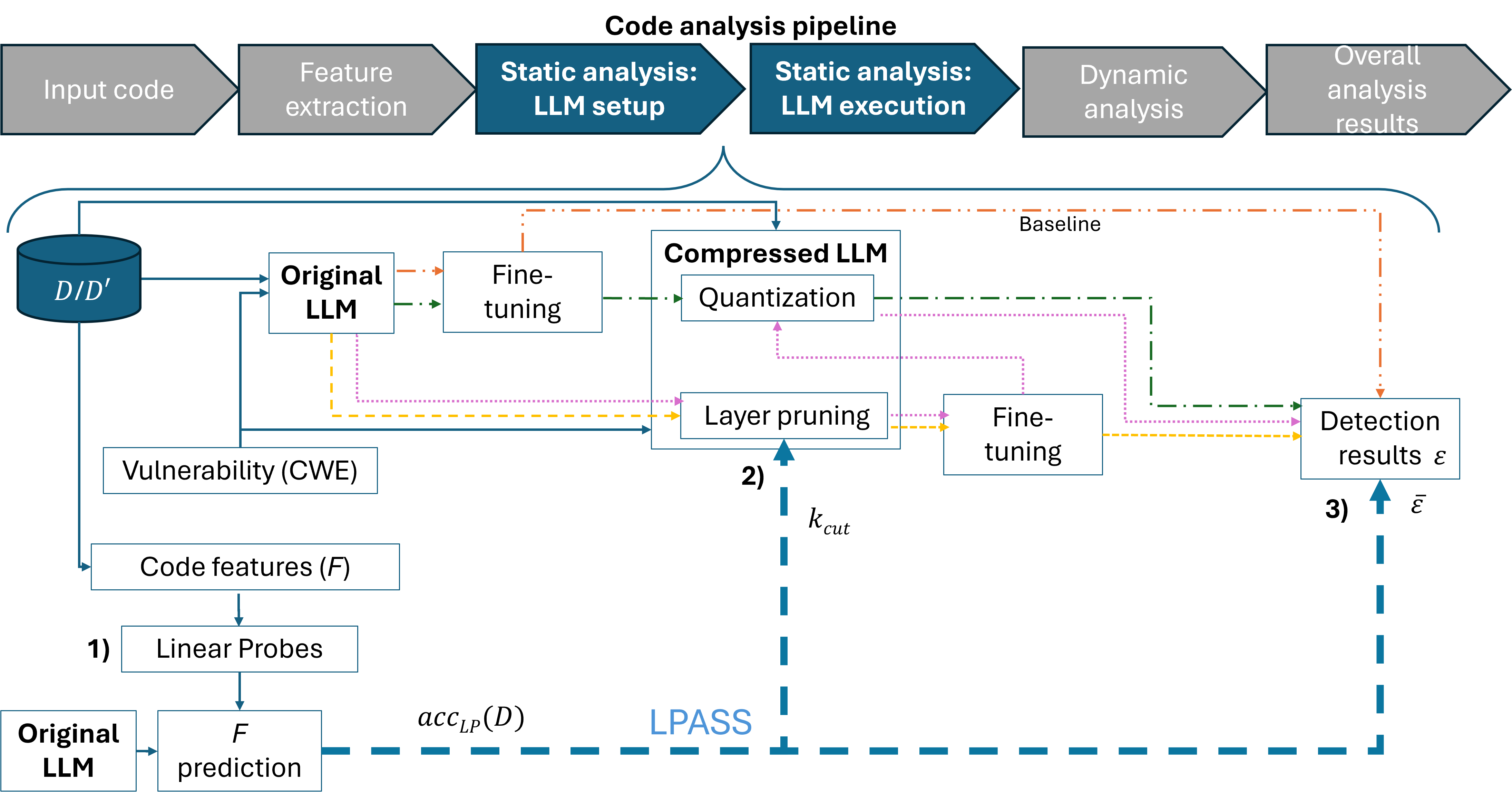}
    \caption{$LPASS$ overview. Steps of $LPASS$ are numbered and colored flows are used to assess the approach when using LLMs to detect vulnerabilities. The orange flow (dot-dot-dash) computes the baseline results, whereas the green (dot-dash), yellow (dashes) and pink (dots) flows refer to compressed LLMs, that is after applying quantization, layer pruning or both techniques at the same time, respectively.
    }
    \label{fig:overview}
    \vspace{-0.5cm}
\end{figure}

\subsection{Overview \& Use case}
\label{sec:overview}
$LPASS$ is framed in a typical vulnerability detection pipeline, which includes both static and dynamic analyses as part of a mobile application security testing\footnote{https://mobile-security.gitbook.io/mobile-security-testing-guide/overview/0x04b-mobile-app-security-testing} as the one in use by Google or Apple in their application store. Thus, codes are subject to a static and dynamic analysis to detect if there are any vulnerabilities (Figure \ref{fig:overview} upper part). The focus of $LPASS$ is on the static analysis phase, that is supposed to be carried out by means of LLMs. 

%{\color{red} Nicolas : fine-tuning does not appear on Fig 1. Possible to add it to the fig ?}

%{\color{red} Nicolas : Same notations should we used here and in section 4.1 for the set of datasets (noted D in 4.1, could be noted $\mathcal D$) and the set of features (noted CF, could be noted F).}

%{\color{red} JF : Insist on that LPs are computed before fine-tuning!}

The aim of $LPASS$ is to help on deciding whether investing resources for fine-tuning and compressing a LLM are worthy. In this regard, the user of $LPASS$ is only requested to extract simple code features $F$ from the samples at stake and compute the LPs. Both tasks involve negligible computational efforts. The remaining values needed for making the decision are the ones provided as a result of our work, as explained later.

Resorting to the selected code features $F$ is essential for a real-world usage of $LPASS$. The user does not know whether the samples are vulnerable or not -- it is the actual purpose of the LLM at stake. However, these code features $F$ can be extracted very easily without the need of any LLM. Therefore, features $F$ can be regarded as \textit{proxies} for the presence or absence of vulnerabilities in the code. Such a feature extraction is natural in a vulnerability detection pipeline, as many of them (e.g., lines of code or the control flow graph) are routinely used. 

The proposed approach is depicted in Figure \ref{fig:overview}. Firstly, for each dataset (see Section \ref{Datasets}),  LPs are trained over the internal activations after each layer of the LLMs, using the code samples as input. The goal is to predict the selected code features $F$ using these activations. The selection of which features $F$ to consider is a key aspect of $LPASS$. Secondly, the accuracy of LPs on each dataset $D$ ($acc_{LP}(D)$) is used to set the cut-off point when compressing the LLM using layer pruning, {\color{black}where all layers beyond the cut-off layer $k_{cut}$} are removed from the LLM. Thirdly, LPs are also used for estimating the vulnerability detection effectiveness ($\overline{\mathcal{E}(D)}$) without the need of fine-tuning or executing any LLM. To assess the quality of the estimations, vulnerabilities are detected ($\mathcal{E}(D)$) using the original LLMs to compute baseline results (orange arrow), as well as in compressed LLMs, that is after applying quantization, layer pruning or both techniques at the same time (green, yellow and pink arrows, respectively). 

It must be noted that in order to validate the quality of the provided estimations, in this paper we carry out both fine-tuning and compression techniques on two representative LLMs. %However, in a real use case, our results are intended to avoid practitioners invest these efforts. 

\subsection{Goals}
\label{sec:goals}
$LPASS$ aims to meet the following goals:
\begin{itemize}
\item \textbf{Early assessment.} It must be possible to ascertain whether the LLM will be effective or not in vulnerability detection while spending a reduced amount of computational efforts. 
    \item \textbf{Vulnerability detection.} $LPASS$-built LLMs must suitably detect vulnerable codes, identifying which vulnerability is present in a given piece of code.
    \item \textbf{Time efficiency.} $LPASS$-built LLMs must minimize the time needed for training and inference.
    \item \textbf{Memory efficiency.} $LPASS$-built LLMs must reduce the memory requirements.
\end{itemize}

As it can be seen, the first goal refers to $LPASS$ itself, whereas the remaining ones are the expected results of its use in LLMs. It must be noted that imposing restrictions on time and memory are aligned to achieving energy savings. 

\section{Description of $LPASS$}
\label{sec:core}
This section describes the proposed method. For the sake of clarity, the two main steps are presented separately -- the way to guide layer pruning using LPs (Section \ref{sec:method}) and using LPs to estimate the post-fine-tuning and post-compression performance in terms of effectiveness, time and model size (Section \ref{sec:estimatemethod}).

\subsection{Layer pruning leveraging Linear Probes}
\label{sec:method}
\begin{figure}
    \centering
    \includegraphics[width=0.4\textwidth]{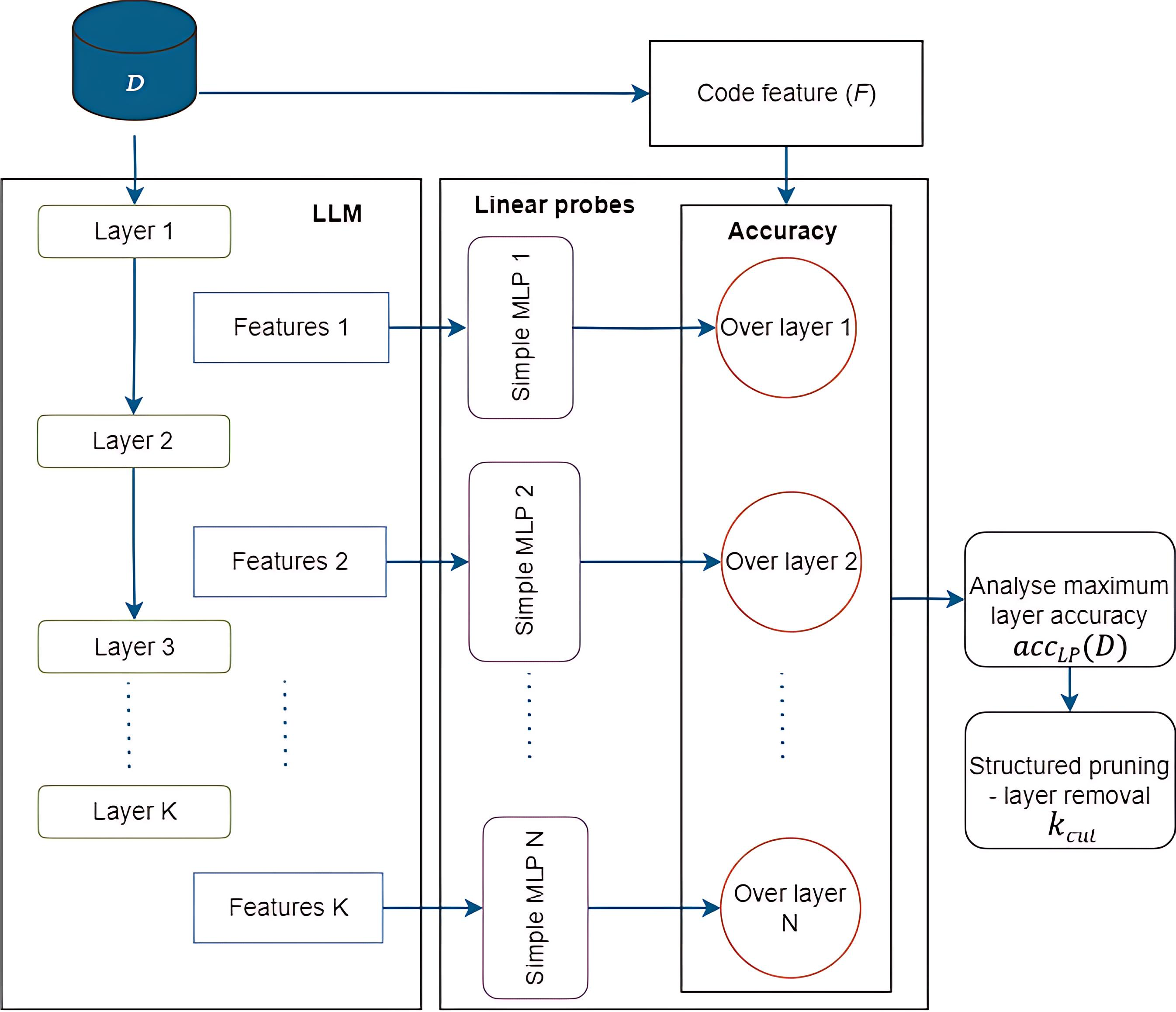}
    \caption{{\color{black}Linear probes used for determining $k_{cut}$. Code features $F$ are the target of the prediction, which is based using the LLM's internal activations per layer. The layer in which the best prediction is made is selected as the cut-off point $k_{cut}$}}
    \label{fig:LP}
    \vspace{-0.5cm}
\end{figure}

%\subsubsection{Layer pruning}
%\label{LP_spruning}
%{\color{red} Nicolas : I think that CC and HD should be given as examples of possible features (and they are actually used in the performance validation section), but not here as part of the general description of the solution. What do you think ?} -- the problem is that they are critical to explain several details

In this proposal, LPs, depicted in Figure \ref{fig:LP}, are implemented with a Multi-Layer Perceptron (MLP) applied at each LLM layer $k$. In particular, the internal activations $H_k$ are collected at each layer $k$ and they are used as input to a MLP which predicts the selected code features $F$.

Once having classes for MLPs, linear probes are computed and their result is the accuracy ($acc$) per layer $k$ and CWE, though the average $acc$ of all CWE per $k$ ($avg(acc_k)$) is determined. Finally, the calculus of the $loss$ of $avg(acc_k)$ allows identifying which is the cutting $k$ ($k_{cut}$) to remove upper layers and enforce structured pruning. The $loss$ refers to the amount of accuracy lost in each $k$ and it is calculated through the difference between the maximum $avg(acc_{k})$, at a particular $k$, and the $avg(acc_k) \ \forall k \in K$, Equation \ref{loss} %{\color{red} JF: Nicolas please help here}: 

%\begin{equation}
%loss_k= (avg(acc_{k_i}) - \max{i \in k} avg(acc_k)) \forall k \in K
 %   \label{loss}
%\end{equation}

%{\color{red} here is a try, but we need to discuss:}
\begin{equation}
\text{loss}_k= \max_{i \in k} ( \text{avg}(\text{acc}_{i})) - \text{avg}(\text{acc}_k) 
    \label{loss}
\end{equation}

This is carried out for all datasets (noted $\mathcal D = \{D\}$) and code features $F$, and $k_{cut}$ is set as the $k$ which minimizes the sum of $loss$, in absolute value, for each $k$, Equation \ref{kcut}.

% OLD:
%\begin{equation} k_{cut} = k \ such \ that \ min( (\sum_{x=1}^{\#Dx\#CF}(|loss_k|) )\ \forall \ k \ \in \ K )    \label{kcut} \end{equation}
\begin{equation} k_{\text{cut}} = \arg \min_{k \in K} \left( \sum_{||\mathcal D|| \times ||F||} |\text{loss}_k| \right) 
\label{kcut} 
\end{equation}

where $||X||$ denotes the cardinality of a set $X$ and $|x|$ the absolute value of $x$.

%\subsubsection{Quantization}

\subsection{Estimating performance for vulnerability detection}
\label{sec:estimatemethod}
The same LPs used for layer pruning are now used for estimating the performance of the LLM to detect vulnerabilities in case compression or fine-tuning techniques were applied. Let $\mathcal{E}(D)$ be the set of effectiveness metrics of a LLM at stake, which contains  precision, recall and F1-score for a dataset $D$. The overall estimation process bases on computing $\overline{\mathcal{E}} \approx \mathcal{E} $ as shown in Equation \ref{estimationa} for each of these metrics. 

\begin{equation}
\overline{\mathcal{E}(D)}= acc_{LP}(D) + \beta
\label{estimationa}
\end{equation}

 The goal is to determine which value $\beta$ should be added  to the result of LPs on $D$, namely their accuracy ($acc_{LP}(D)$), to obtain the estimate. %Interestingly, this estimation can be carried out in each LLM, either compressed or just in the baseline setting.
Note that $acc_{LP}(D) \in [0,1]$, and $\beta \in [0,1]$ as explained later. Thus, $\overline{\mathcal{E}(D)}$ must be post-normalized to fit into $[0,1]$ as well.
 
 \subsubsection{Computing $\beta$} $\beta$ is the value provided within this proposal to be used in the equation above. It is computed leveraging a knowledge base formed by datasets $D' \neq D$, as shown in Equation \ref{estimationbeta}.

 \begin{equation}
\beta = \mathcal{E}(D') - acc_{LP}(D')
\label{estimationbeta}
\end{equation}

In order to get $\beta$, it is necessary to obtain $\mathcal{E}(D')$, that is, the real performance result for the LLM on the dataset $D'$. However, note that users of $LPASS$ will not need to perform this operation -- we later show that our results for $\beta$ generalize for different datasets.

 \subsubsection{Validating $\overline{\mathcal{E}(D)}$} To assess the approach, an \textit{Estimation Error} ($Err$) is computed to determine the error incurred, as shown in Equation \ref{EE}. 

 \begin{equation}
Err= \overline{\mathcal{E}(D)} - \mathcal{E}(D)
\label{EE}
\end{equation}

In our approach, we select different datasets (as introduced later) and we apply a leave-one-out cross validation. This ensures the generalization of our results.

\section{Evaluation}
\label{sec:assess}
This section assesses $LPASS$. For this purpose, the LLMs and data at stake are introduced in Sections \ref{LLMs} and \ref{Datasets}, respectively. The selected code features $F$ are described in Section \ref{codeF}. Sections \ref{sec:expset} and \ref{sec:metrics} describe the experimental settings and metrics. Results are presented in Section \ref{sec:results}. Lastly, Section \ref{Discussion} discusses the results and the limitations of the approach.

\subsection{LLMs}
\label{LLMs}
{\color{black} A couple of large language models (LLMs) are selected. On one hand, the BERT model is chosen due to its widespread use \cite{zhou2024large}, and its prominence in related works such as \cite{chen2023diversevul,shi2024greening,hanif2022vulberta}. The large variant of BERT is used, featuring 334 million parameters and 24 layers.} %As observed by Li et al. \cite{li2020train} larger models are more robust to pruning and compression {\color{red}LUIS<---this sentence is not valuable to me}. 
On the other hand, Gemma model is chosen for being one of the latest and most recently used LLMs \cite{team2024gemma}, configured, due to resource limitations, with 2 billions of parameters and 18 layers. %Besides, lighter training methods has to be used to be able to fine-tune this model with a consumer user GPU.  For so, Galore methodology is employed \cite{zhao2024galore}.

All LLMs base on a fundamental unit, the token, which is the minimum processed information. In the case of BERT, 512 tokens are the maximum applied, while in Gemma this number can increase up to 8,192. Nonetheless, our resource constraints force the use of Gemma with up to 1,024 tokens.

\subsection{Datasets and Vulnerabilities}
\label{Datasets}
Three C/C++ datasets, are selected for being well-known and for facilitating a multi-class classification. Indeed, they are chosen among those used in the state of art (cf. Section \ref{RW}).  DiverseVul \cite{chen2023diversevul} is composed of extracted commits from vulnerable and non-vulnerable functions covering more than 295 projects after crawling security issue websites. Big-Vul \cite{fan2020ac} is a vulnerability dataset from 348 open source Github projects, where
vulnerability-related code commits and extracted
relevant code changes are collected. PrimeVul \cite{ding2024vulnerability} is a dataset of benign and vulnerable functions composed of a merging of four well-known vulnerability datasets, including both previously mentioned. Note that based on related work (Section \ref{RW}), Devign and Draper datasets were also considered but discarded since they do not provide information about the CWE, just vulnerable or not. Similarly, CVE-fixes and SARD were analysed and discarded because the amount of samples per CWE  were smaller (3k maximum) than the ones of the chosen datasets.

Among all existing vulnerabilities, the Top 25 most dangerous CWE \footnote{https://cwe.mitre.org/top25/ , last access September 2024} are considered herein. As not all CWEs appear in all datasets, we select the 10 CWEs with the highest number of samples per dataset. Table \ref{tab:Datasets} summarizes the number of samples per dataset and selected CWE, most of them common for all datasets. Hence, a total of 12 CWEs are at stake. It is noteworthy that 1,024 token size allows increasing the sample set, thus leading to 480k samples. Samples beyond 512 or 1,024 tokens on each of the cases have been removed.

% Table generated by Excel2LaTeX from sheet 'Datasets'
\begin{table}[htbp]
  \centering
  \caption{Samples per dataset of the 10 most represented CWEs within MITRE Top 25}
  	\scalebox{0.55}{
    \begin{tabular}{p{4em}|l|r|r|r||r|r|r|}
\cline{3-8}    \multicolumn{1}{r}{} &       & \multicolumn{3}{c||}{\textbf{512 Tokens}} & \multicolumn{3}{c|}{\textbf{1,024 Tokens}} \\
\cline{3-8}      \multicolumn{1}{r}{}     &       & \multicolumn{1}{c|}{\textbf{DiverseVul}} & \multicolumn{1}{c|}{\textbf{Big-Vul}} & \multicolumn{1}{c||}{\textbf{PrimeVul}} & \multicolumn{1}{c|}{\textbf{DiverseVul}} & \multicolumn{1}{c|}{\textbf{Big-Vul}} & \multicolumn{1}{c|}{\textbf{PrimeVul}} \\
\cline{2-8}    \multicolumn{1}{r|}{} & \textbf{CWE} & \multicolumn{3}{c||}{\textbf{\# samples}} & \multicolumn{3}{c|}{\textbf{\# samples}} \\
    \hline
    \multicolumn{1}{|p{4em}|}{\multirow{8}{*}{\textbf{Common}}} & 20    & 22,197 & 15,326 & 25,799 & 33,491 & 25,799 & 25,596 \\
\cline{2-8}    \multicolumn{1}{|c|}{} & 119   & 19,853 & 18,321 & 22,575 & 32,013 & 22,575 & 17,216 \\
\cline{2-8}    \multicolumn{1}{|c|}{} & 125   & 23,675 & 4,774 & 2,498 & 32,951 & 6,206 & 19,377 \\
\cline{2-8}    \multicolumn{1}{|c|}{} & 190   & 8,905 & 2,632 & \multicolumn{1}{r||}{3,900} & 29,395 & 3,262 & 7,524 \\
\cline{2-8}    \multicolumn{1}{|c|}{} & 362   & 7,729 & 4,363 & 6,206 & 11,493 & 5,478 & 6,882 \\
\cline{2-8}    \multicolumn{1}{|c|}{} & 416   & 23,691 & 7,428 & 8,993 & 32,951 & 8,993 & 16,593 \\
\cline{2-8}    \multicolumn{1}{|c|}{} & 476   & 16,691 & 2,943 & 5,478 & 27,981 & 3,900 & 12,781 \\
\cline{2-8}    \multicolumn{1}{|c|}{} & 787   & 19,241 & 2,020 & 3,262 & 29,395 & 2,498 & 27,199 \\
    \hline
    \multicolumn{1}{|p{4em}|}{\multirow{4}{*}{\textbf{Different}}} & 22    & 2,409 & \multicolumn{1}{c|}{-} & \multicolumn{1}{c||}{-} & 3,693 & \multicolumn{1}{c|}{-} & \multicolumn{1}{c|}{-} \\
\cline{2-8}    \multicolumn{1}{|c|}{} & 78    & 2,233 & \multicolumn{1}{c|}{-} & 875   & 3,051 & \multicolumn{1}{c|}{-} & 29,066 \\
\cline{2-8}    \multicolumn{1}{|c|}{} & 79    & \multicolumn{1}{c|}{-} & 699   & 857   & \multicolumn{1}{c|}{-} & 857   & 653 \\
\cline{2-8}    \multicolumn{1}{|c|}{} & 269   & \multicolumn{1}{c|}{-} & 639   & \multicolumn{1}{c||}{-} & \multicolumn{1}{c|}{-} & 875   & \multicolumn{1}{c|}{-} \\
    \hline
          & \textbf{Total} & 146,624 & 59,145 & 76,543 & 236,414 & 80,443 & 162,887 \\
\cline{2-8}    \end{tabular}%
}
   \label{tab:Datasets}%
\end{table}%

\subsection{Code features $F$}
\label{codeF}
Cyclomatic Complexity (CC) and Halstead Difficulty (HD) of code samples are the chosen code features $F$. There are lots of $F$ that could be predicted but looking for capturing code complexity \cite{herraiz2010beyond} and considering the state of the art in vulnerability detection \cite{zagane2020deep, lomio2022just}, both CC and HD were chosen for their ability to capture the structural information of the code. CC is a quantitative measure of the number of linearly independent paths in the code \cite{ebert2016cyclomatic}, while HD measures the diversity of operands present in the code\footnote{https://product-help.schneider-electric.com/Machine\%20Expert/V2.0/en/CodeAnly/CodeAnly/D-SE-0095969.html, last accessed September 2024}. Accurately predicting these features would indicate that the LLMs effectively encode the structural information of the code, offering a more abstract and complex understanding than simply predicting basic metrics like the number of lines or tokens. These features are particularly valuable because they are challenging to predict, requiring the LLMs to deeply understand and analyze the code, unlike simpler metrics such as line counts.

CC and HD are integer and float numbers respectively, from 0 to infinite and thus, the number of classes of the MLP has to be determined. To do so, CC and HD are computed for all samples in each dataset and after the analysis of their distribution, the number of classes is set such that the vast majority of samples of each dataset (in our case, we opted for 85\%) are included.

\subsection{Experimental settings}
\label{sec:expset}
This section outlines the training settings for the chosen models. Our experimental materials are publicly released\footnote{\textit{A reduced version is published until acceptance; https://github.com/Luisibear98/LPASS-pruning}}. Training was conducted on two NVIDIA consumer GPUs, a RTX 4090 and a RTX 4080, using the Pytorch framework and the Hugging Face library \footnote{https://huggingface.co}. For both training and validation, all datasets were split in an 80\%-20\% random distribution to accommodate the imbalance of data. Each CWE is limited to 5,000 samples for training because, after a trial and error process, such value allows the execution of both models, while lower ones do not work in both of them. To ensure balanced CWE classes, if a CWE does not have enough samples, oversampling is carried out copying samples until getting to the set limit \cite{mohammed2020machine}. Otherwise, undersampling is enforced removing samples \cite{mohammed2020machine}, though setting aside 20\% for the validation set. Besides, computations are repeated 3 times per model and pruning set-up and results present the average of all executions.

BERT model is fine-tuned for training based on \cite{chen2023diversevul}, specifically a learning rate of 2e-5 is applied, using the Adam optimizer \cite{kingma2014adam} over 10 epochs. For Gemma, Galore low-rank adaptation training \cite{zhao2024galore} was adopted due to resource constraints, using per layer-weights update implementation. Since higher rank requires more GPU memory,  when fine-tuning the entire model, the rank is limited to 256 but fine-tuning pruned models it is set to 1,024.

In what comes to the batch size, our preliminary tests show that small sizes
do not affect the accuracy while harm performance. Thus,
the batch size is set to the maximum capacity per GPU.
Finally, concerning the class 'non vulnerable', in DiverseVul it corresponds to samples marked as non-vulnerable, in PrimeVul and big-vul refers to samples tagged as 'None' in the CWE-ID field.

\subsection{Metrics}
\label{sec:metrics}

\textit{Effectiveness metrics} used to measure the model's capabilities in a multi-class classification problem include Accuracy, F1 Score, Precision, and Recall. These metrics are reported in two ways. First, they assess the model’s overall ability to discern among multiple classes, reflecting its pure capacity in a multi-class context. Second, they evaluate the model in a binary fashion, determining its ability to differentiate between code with a CWE and code without a CWE, thus considered no vulnerable. Indeed, in this latter case, given the  class imbalance (5,000 non vulnerable vs 50,000 vulnerable), F1 is the computed metric. 
\begin{comment}
    
{\color{red}due to space limitations, I would remove equations, they are really well-known}

In this manner, Accuracy measures the overall correctness of the model across all classes, Equation \ref{acc}.

\begin{equation}
\text{Accuracy} = \frac{\text{Correct Predictions}}{\text{Total Number of Predictions}}
\label{acc}    
\end{equation}

Precision is reported as the ratio of correctly predicted positive observations (TP) to the total predicted positives for that class, Equation \ref{precision}:
\begin{equation}
\text{Precision} = \frac{\text{True Positives}}{\text{True Positives} + \text{False Positives}}
\label{precision}    
\end{equation}

Recall references to the ratio of correctly predicted positive observations to all observations in the actual class, Equation \ref{recall}:

\begin{equation}
\text{Recall} = \frac{\text{True Positives}}{\text{True Positives} + \text{False Negatives}}
\label{recall}    
\end{equation}

Finally, F1 Score is measured as the weighted average of Precision and Recall, Equation \ref{f1}:

\begin{equation}
F1 = 2 \cdot \frac{\text{Precision} \cdot \text{Recall}}{\text{Precision} + \text{Recall}}
\label{f1}  
\end{equation}

\end{comment}
Additionally, \textit{performance metrics} refer to time and memory ones. The former ones corresponds to training time of the model and the inference one required for each sample prediction. By contrast, memory metrics refer to the minimum amount of GPU memory required for training and inference the model setting a batch size of 1, the number of effective parameters and the model size. %{\color{red} LUIS: Revisar si la explicacion de inference es correcta y explicar effective parameters}

%All above metrics are expressed in percentages (\%) based on the difference with baseline results, that is those of the original models. 

\subsection{Results}
\label{sec:results}
 This section presents results of the enforcement of structured pruning applying LP (Section \ref{LPruningR}), the detection of vulnerability (Section \ref{VulnDetec}), the effects of the detection process in time and memory efficiency (Section \ref{TMefficiency}) and the early assessment of the model effectiveness after fine-tuning and compression (Section \ref{EarlyAssess}).

 \textbf{Understanding Tables \ref{tab:resultsVulDetectPruning} and \ref{tab:resultsVulDetectQuantPruning}.} These tables summarize the impact of layer pruning using the cut-off point (Table \ref{tab:resultsVulDetectPruning}) and also quantization (Table \ref{tab:resultsVulDetectQuantPruning}) on effectiveness, time, model memory and size. For clarity, baseline results are shown in gray in Table \ref{tab:resultsVulDetectPruning}. The results of the models after applying these compression techniques are then presented as the \textit{difference (in \%) between the baseline and each compressed version}. Thus, negative values in effectiveness are preferred -- this means that the compressed model outperforms the baseline. On the contrary, positive values in time and model memory/size are desired -- the compressed model would then be faster and smaller. 

\begin{table*}[htbp]
  \centering
  \caption{Vulnerability detection, time and memory results - Baseline (in gray) \& Layer pruning (\% vs baseline). For effectiveness columns, a negative value refers to an increment over the baseline.  For time and model, a positive values means an improvement over the baseline.
  }
   \scalebox{0.55}{
    \begin{tabular}{|l|c|r|r|r|r|r|r|r|r|r|r|r|}
\cline{3-13}    \multicolumn{1}{r}{} &       & \multicolumn{5}{c|}{\textbf{Effectiveness}} & \multicolumn{2}{c|}{\multirow{2}[4]{*}{\textbf{Time}}} & \multicolumn{4}{c|}{\multirow{2}[4]{*}{\textbf{Model}}} \\
\cline{3-7}    \multicolumn{1}{r}{} &       & \multicolumn{4}{c|}{\textbf{Multi-class detection}} & \multicolumn{1}{p{4.5em}|}{\textbf{Binary detection}} & \multicolumn{2}{c|}{} & \multicolumn{4}{c|}{} \\
    \hline
    \multicolumn{1}{|p{2.335em}|}{\textbf{Rank}} & \multicolumn{1}{p{4.165em}|}{\textbf{Removed layers}} & \multicolumn{1}{p{4.335em}|}{\textbf{Accuracy}} & \multicolumn{1}{p{4.055em}|}{\textbf{F1}} & \multicolumn{1}{p{4.11em}|}{\textbf{Precision}} & \multicolumn{1}{p{4.11em}|}{\textbf{Recall}} & \multicolumn{1}{p{4.5em}|}{\textbf{F1}} & \multicolumn{1}{p{4.28em}|}{\textbf{Training time}} & \multicolumn{1}{p{4.665em}|}{\textbf{Inference time}} & \multicolumn{1}{p{4.72em}|}{\textbf{Memory GPU- train}} & \multicolumn{1}{p{4.39em}|}{\textbf{Memory GPU-inference}} & \multicolumn{1}{p{7.165em}|}{\textbf{Effective Parameters}} & \multicolumn{1}{p{6.11em}|}{\textbf{Model size}} \\
    \hline
    \multicolumn{13}{|c|}{\textbf{Gemma (18 layers)}} \\
    \hline
    \multicolumn{13}{|c|}{\textbf{DiverseVul}} \\
    \hline
    \rowcolor[rgb]{ .949,  .949,  .949} \multicolumn{1}{|c|}{256} & 0     & \multicolumn{1}{c|}{73.40\%} & \multicolumn{1}{c|}{74.90\%} & \multicolumn{1}{c|}{75.30\%} & \multicolumn{1}{c|}{75.00\%} & \multicolumn{1}{c|}{95.90\%} & \multicolumn{1}{p{4.28em}|}{3h 10 min} & \multicolumn{1}{p{4.665em}|}{0.018s} & \multicolumn{1}{p{4.72em}|}{8.96GB} & \multicolumn{1}{p{4.39em}|}{5.76GB} & \multicolumn{1}{c|}{2506172416} & \multicolumn{1}{p{6.11em}|}{4770.15MB} \\
    \hline
    \multicolumn{1}{|c|}{256} & 13    & \multicolumn{1}{c|}{0.8} & \multicolumn{1}{c|}{0.53} & \multicolumn{1}{c|}{0.43} & \multicolumn{1}{c|}{0.87} & \multicolumn{1}{c|}{-0.06} & \multicolumn{1}{c|}{30.18} & \multicolumn{1}{c|}{\multirow{2}[4]{*}{78}} & \multicolumn{1}{c|}{41.07} & \multicolumn{1}{c|}{\multirow{2}[4]{*}{48.09}} & \multicolumn{1}{c|}{\multirow{2}[4]{*}{52.72}} & \multicolumn{1}{c|}{\multirow{2}[4]{*}{57.02}} \\
\cline{1-8}\cline{10-10}    \multicolumn{1}{|c|}{1024} & 13    & \multicolumn{1}{c|}{-2.19} & \multicolumn{1}{c|}{-2.16} & \multicolumn{1}{c|}{-1.9} & \multicolumn{1}{c|}{-2.16} & \multicolumn{1}{c|}{-0.13} & \multicolumn{1}{c|}{52.63} &       & \multicolumn{1}{c|}{34.98} &       &       &  \\
    \hline
    \multicolumn{13}{|c|}{\textbf{PrimeVul}} \\
    \hline
    \rowcolor[rgb]{ .949,  .949,  .949} \multicolumn{1}{|c|}{256} & 0     & \multicolumn{1}{c|}{88.10\%} & \multicolumn{1}{c|}{89.00\%} & \multicolumn{1}{c|}{88.80\%} & \multicolumn{1}{c|}{88.10\%} & \multicolumn{1}{c|}{99.50\%} & \multicolumn{1}{p{4.28em}|}{3h 31m 1s} & \multicolumn{1}{p{4.665em}|}{0.013s} & \multicolumn{1}{p{4.72em}|}{8.96GB} & \multicolumn{1}{p{4.39em}|}{5.46GB} & \multicolumn{1}{c|}{2506172416} & \multicolumn{1}{p{6.11em}|}{4770.15MB} \\
    \hline
    \multicolumn{1}{|c|}{256} & 13    & \multicolumn{1}{c|}{\textbf{1}} & \multicolumn{1}{c|}{1.6} & \multicolumn{1}{c|}{0.2} & \multicolumn{1}{c|}{1.61} & \multicolumn{1}{c|}{-0.06} & \multicolumn{1}{c|}{70.62} & \multicolumn{1}{c|}{\multirow{2}[4]{*}{76.15}} & 41.07 & \multicolumn{1}{c|}{\multirow{2}[4]{*}{49.08}} & \multicolumn{1}{c|}{\multirow{2}[4]{*}{52.72}} & \multicolumn{1}{c|}{\multirow{2}[4]{*}{57.02}} \\
\cline{1-8}\cline{10-10}    \multicolumn{1}{|c|}{1024} & 13    & \multicolumn{1}{c|}{1.2} & \multicolumn{1}{c|}{1.5} & \multicolumn{1}{c|}{0.5} & \multicolumn{1}{c|}{1.17} & \multicolumn{1}{c|}{-0.02} & \multicolumn{1}{c|}{71.36} &       & 34.98 &       &       &  \\
    \hline
    \multicolumn{13}{|c|}{\textbf{Big-Vul}} \\
    \hline
    \rowcolor[rgb]{ .949,  .949,  .949} \multicolumn{1}{|c|}{256} & 0     & \multicolumn{1}{c|}{77.00\%} & \multicolumn{1}{c|}{79.30\%} & \multicolumn{1}{c|}{78.80\%} & \multicolumn{1}{c|}{79.90\%} & \multicolumn{1}{c|}{95.30\%} & \multicolumn{1}{p{4.28em}|}{3h 46m 16s} & \multicolumn{1}{p{4.665em}|}{0.013s} & \multicolumn{1}{p{4.72em}|}{8.96GB} & \multicolumn{1}{p{4.39em}|}{5.46GB} &       & \multicolumn{1}{p{6.11em}|}{4770.15MB} \\
    \hline
    \multicolumn{1}{|c|}{256} & 13    & \multicolumn{1}{c|}{-0.51} & \multicolumn{1}{c|}{-1.42} & \multicolumn{1}{c|}{-2.04} & \multicolumn{1}{c|}{-0.83} & \multicolumn{1}{c|}{0.43} & \multicolumn{1}{c|}{65.54} & \multicolumn{1}{c|}{\multirow{2}[4]{*}{76.15}} & 41.07 & \multicolumn{1}{c|}{\multirow{2}[4]{*}{49.08}} & \multicolumn{1}{c|}{\multirow{2}[4]{*}{52.72}} & \multicolumn{1}{c|}{\multirow{2}[4]{*}{57.02}} \\
\cline{1-8}\cline{10-10}    \multicolumn{1}{|c|}{1024} & 13    & \multicolumn{1}{c|}{-2.69} & \multicolumn{1}{c|}{-3.48} & \multicolumn{1}{c|}{-4.59} & \multicolumn{1}{c|}{-2.63} & \multicolumn{1}{c|}{-0.78} & \multicolumn{1}{c|}{58.81} &       & 34.98 &       &       &  \\
    \hline
    \hline
    \multicolumn{13}{|c|}{\textbf{BERT (24 layers)}} \\
    \hline
    \multicolumn{13}{|c|}{\textbf{DiverseVul}} \\
    \hline
    \rowcolor[rgb]{ .949,  .949,  .949} -     & 0     & 78.20\% & 79.60\% & 79.50\% & 79.60\% & 96.10\% & \multicolumn{1}{p{4.28em}|}{2h 48m 49s} & \multicolumn{1}{p{4.665em}|}{0.0056s} & \multicolumn{1}{p{4.72em}|}{7.21 GB} & \multicolumn{1}{p{4.39em}|}{1.726 GB} & 335153163 & \multicolumn{1}{p{6.11em}|}{1278.46 MB} \\
    \hline
    -     & 8     & 0.48  & 0.92  & 0.44  & 1.14  & 0.28  & 47.47 & 32.14 & 27.77 & 15.41 & 35.7  & 26.32 \\
    \hline
    \multicolumn{13}{|c|}{\textbf{PrimeVul}} \\
    \hline
    \rowcolor[rgb]{ .949,  .949,  .949} -     & 0     & 86.10\% & 86.90\% & 87.10\% & 86.60\% & 99.20\% & \multicolumn{1}{p{4.28em}|}{2h 46m 26s} & \multicolumn{1}{p{4.665em}|}{0.0056s} & \multicolumn{1}{p{4.72em}|}{7.21 GB} & \multicolumn{1}{p{4.39em}|}{1.726 GB} & 335153163 & \multicolumn{1}{p{6.11em}|}{1278.46 MB} \\
    \hline
    -     & 8     & 0.2   & 0.2   & 0.52  & -0.27 & 0.21  & 68.54 & 32.14 & 27.77 & 15.41 & 26.31 & 26.32 \\
    \hline
    \multicolumn{13}{|c|}{\textbf{Big-Vul}} \\
    \hline
    \rowcolor[rgb]{ .949,  .949,  .949} -     & 0     & 82.40\% & 85.10\% & 85.40\% & 84.70\% & 95.90\% & \multicolumn{1}{p{4.28em}|}{2h 38m 47s} & \multicolumn{1}{p{4.665em}|}{0.0056s} & \multicolumn{1}{p{4.72em}|}{7.21 GB} & \multicolumn{1}{p{4.39em}|}{1.726 GB} & 335153163 & \multicolumn{1}{p{6.11em}|}{1278.46 MB} \\
    \hline
    -     & 8     & 0.34  & 0.45  & 0.57  & 1.9   & 0.06  & 68.91 & 32.14 & 27.77 & 15.41 & 26.31 & 26.32 \\
    \hline
\end{tabular}%
	}
  \label{tab:resultsVulDetectPruning}%
\end{table*}%

\begin{table*}[htbp]
  \centering
  \caption{Vulnerability detection, time and memory results - Quantization \& Layer pruning (\% vs baseline shown in Table \ref{tab:resultsVulDetectPruning})}
     \scalebox{0.7}{
    \begin{tabular}{|p{5.665em}|c|r|r|r|r|r|r|r|r|r|}
\cline{3-11}    \multicolumn{1}{r}{} &       & \multicolumn{5}{c|}{\textbf{Effectiveness}} & \multicolumn{1}{c|}{\multirow{2}[4]{*}{\textbf{Time}}} & \multicolumn{3}{c|}{\multirow{2}[4]{*}{\textbf{Model}}} \\
\cline{3-7}    \multicolumn{1}{r}{} &       & \multicolumn{4}{c|}{\textbf{Multi-class detection}} & \multicolumn{1}{p{4.445em}|}{\textbf{Binary detection}} &       & \multicolumn{3}{c|}{} \\
    \hline
    \textbf{Quantization} & \multicolumn{1}{p{4.665em}|}{\textbf{Removed layers}} & \multicolumn{1}{p{4.165em}|}{\textbf{Accuracy}} & \multicolumn{1}{p{2.39em}|}{\textbf{F1}} & \multicolumn{1}{p{4.165em}|}{\textbf{Precision}} & \multicolumn{1}{p{2.72em}|}{\textbf{Recall}} & \multicolumn{1}{p{4.445em}|}{\textbf{F1}} & \multicolumn{1}{p{4.11em}|}{\textbf{Inference time}} & \multicolumn{1}{p{4.11em}|}{\textbf{Memory GPU-inference (GB}} & \multicolumn{1}{p{5.39em}|}{\textbf{Effective Parameters}} & \multicolumn{1}{p{5.39em}|}{\textbf{Model size}} \\
    \hline
    \multicolumn{11}{|c|}{\textbf{Gemma (18 layers/ 256 rank)}} \\
    \hline
    \multicolumn{11}{|c|}{\textbf{DiverseVul}} \\
    \hline
    8 bits & 0     & \multicolumn{1}{c|}{1} & \multicolumn{1}{c|}{1.32} & \multicolumn{1}{c|}{1} & \multicolumn{1}{c|}{1.25} & \multicolumn{1}{c|}{1.43} & \multicolumn{1}{c|}{-763.49} & \multicolumn{1}{c|}{36.46} & 0     & 39.35 \\
    \hline
    4 bits & 0     & \multicolumn{1}{c|}{18.78} & \multicolumn{1}{c|}{19.8} & \multicolumn{1}{c|}{10.65} & \multicolumn{1}{c|}{19.68} & \multicolumn{1}{c|}{2.18} & \multicolumn{1}{c|}{-400.5} & \multicolumn{1}{c|}{43.98} & 39.54 & 56.75 \\
    \hline
    8 bits & 13    & \multicolumn{1}{c|}{1.3} & \multicolumn{1}{c|}{1.48} & \multicolumn{1}{c|}{0.71} & \multicolumn{1}{c|}{1.2} & \multicolumn{1}{c|}{-0.43} & \multicolumn{1}{c|}{-4.5} & \multicolumn{1}{c|}{51.35} & 57.11 & 68.01 \\
    \hline
    4 bits & 13    & \multicolumn{1}{c|}{1.2} & \multicolumn{1}{c|}{1.26} & \multicolumn{1}{c|}{0.6} & \multicolumn{1}{c|}{1.19} & \multicolumn{1}{c|}{-0.4} & \multicolumn{1}{c|}{20.8} & \multicolumn{1}{c|}{57.33} & 68.09 & 72.84 \\
    \hline
    \multicolumn{11}{|c|}{\textbf{PrimeVul}} \\
    \hline
    8 bits & 0     & \multicolumn{1}{c|}{0.17} & \multicolumn{1}{c|}{0.04} & \multicolumn{1}{c|}{0.24} & \multicolumn{1}{c|}{-0.71} & \multicolumn{1}{c|}{-0.04} & \multicolumn{1}{p{4.11em}|}{-2.130.77} & \multicolumn{1}{c|}{32.97} & 0     & 39.35 \\
    \hline
    4 bits & 0     & \multicolumn{1}{c|}{22.38} & \multicolumn{1}{c|}{27.4} & \multicolumn{1}{c|}{15.4} & \multicolumn{1}{c|}{30.85} & \multicolumn{1}{c|}{1.08} & \multicolumn{1}{c|}{-300} & \multicolumn{1}{c|}{40.90} & 39.54 & 56.75 \\
    \hline
    8 bits & 13    & \multicolumn{1}{c|}{1} & \multicolumn{1}{c|}{1.5} & \multicolumn{1}{c|}{2.46} & \multicolumn{1}{c|}{1.11} & \multicolumn{1}{c|}{-0.02} & \multicolumn{1}{c|}{-223.08} & \multicolumn{1}{c|}{48.68} & 57.11 & 68.01 \\
    \hline
    4 bits & 13    & \multicolumn{1}{c|}{4.1} & \multicolumn{1}{c|}{3.1} & \multicolumn{1}{c|}{4.1} & \multicolumn{1}{c|}{1.18} & \multicolumn{1}{c|}{0} & \multicolumn{1}{c|}{-153.85} & \multicolumn{1}{c|}{54.98} & 68.09 & 72.84 \\
    \hline
    \multicolumn{11}{|c|}{\textbf{Big-Vul}} \\
    \hline
    8 bits & 0     & \multicolumn{1}{c|}{0.05} & \multicolumn{1}{c|}{0.77} & \multicolumn{1}{c|}{0.56} & \multicolumn{1}{c|}{0.41} & \multicolumn{1}{c|}{0.42} & \multicolumn{1}{c|}{-846.15} & \multicolumn{1}{c|}{32.97} & 0     & 39.35 \\
    \hline
    4 bits & 0     & \multicolumn{1}{c|}{10.05} & \multicolumn{1}{c|}{9.92} & \multicolumn{1}{c|}{5.46} & \multicolumn{1}{c|}{10.06} & \multicolumn{1}{c|}{1} & \multicolumn{1}{c|}{-420.77} & \multicolumn{1}{c|}{40.90} & 39.54 & 56.75 \\
    \hline
    8 bits & 13    & \multicolumn{1}{c|}{-2.05} & \multicolumn{1}{c|}{-2.13} & \multicolumn{1}{c|}{-2.91} & \multicolumn{1}{c|}{-1.76} & \multicolumn{1}{c|}{0.22} & \multicolumn{1}{c|}{-190.77} & \multicolumn{1}{c|}{48.68} & 57.11 & 68.01 \\
    \hline
    4 bits & 13    & \multicolumn{1}{c|}{-1.25} & \multicolumn{1}{c|}{-1.61} & \multicolumn{1}{c|}{-2.6} & \multicolumn{1}{c|}{-1.23} & \multicolumn{1}{c|}{0.8} & \multicolumn{1}{c|}{-69.23} & \multicolumn{1}{c|}{54.98} & 68.09 & 72.84 \\
    \hline
    \hline
    \multicolumn{11}{|c|}{\textbf{BERT (24 layers)}} \\
    \hline
    \multicolumn{11}{|c|}{\textbf{DiverseVul}} \\
    \hline
    8 bits & 0     & 0.32  & 0.23  & -0.12 & 0.38  & 0.2   & -846.43 & 18.31 & 0     & 72.53 \\
    \hline
    4 bits & 0     & 0.89  & 0.71  & 0.3   & 0.88  & 0.39  & -176.79 & 6.72  & 82.51 & 80.08 \\
    \hline
    8 bits & 8     & 0.59  & 1.28  & 0.58  & 1.27  & 0.38  & -517.86 & 30.48 & 51.11 & 82.48 \\
    \hline
    4 bits & 8     & 1.7   & 1.36  & 1     & 1.51  & 0.33  & -85.71 & 18.89 & 164.13 & 86.73 \\
    \hline
    \multicolumn{11}{|c|}{\textbf{PrimeVul}} \\
    \hline
    8 bits & 0     & 0     & -0.08 & -0.18 & -0.12 & 0.02  & -828.57 & 17.15 & 0     & 72.53 \\
    \hline
    4 bits & 0     & 0.14  & 0.07  & -0.24 & 0.22  & 0.01  & -167.86 & 3.82  & 82.51 & 80.08 \\
    \hline
    8 bits & 8     & 0.5   & 0.47  & 1.1   & 0.53  & 0.11  & -507.14 & 31.63 & 51.11 & 82.48 \\
    \hline
    4 bits & 8     & 1     & 1.2   & 1.1   & 0.8   & 0.11  & -78.57 & 21.21 & 164.13 & 86.73 \\
    \hline
    \multicolumn{11}{|c|}{\textbf{Big-Vul}} \\
    \hline
    8 bits & 0     & 0.26  & 0.18  & 0.03  & 0.3   & -0.12 & -828.57 & 13.67 & 0     & 72.53 \\
    \hline
    4 bits & 0     & 0.95  & 0.58  & 0.34  & 0.78  & -0.05 & -167.86 & 11.94 & 82.51 & 80.08 \\
    \hline
    8 bits & 8     & 0.64  & 1.05  & 1.23  & 0.82  & 0.15  & -525  & 28.56 & 51.11 & 82.48 \\
    \hline
    4 bits & 8     & 0.84  & 1.66  & 1.63  & 1.7   & 0.09  & -87.5 & 24.1  & 164.13 & 86.73 \\
    \hline
    \end{tabular}%
	}
  \label{tab:resultsVulDetectQuantPruning}%
\end{table*}%

%\begin{figure}
 %   \centering
  %  \includegraphics[width=0.4\textwidth]{images/MLP_classes_CC.png}
  %  \caption{Distribution of samples per CC class}
   % \label{fig:classesselection}
%\end{figure}

\subsubsection{Layer pruning}
\label{LPruningR}
This step involves the computation of LP. Firstly, MLP classes are established based on Section \ref{sec:core}. CC and HD are calculated getting the average of all CWE. A similar distribution is identified in all cases. %For illustration,  Figure \ref{fig:classesselection} presents results for CC. 
Considering a coverage of 85\% or more (recall Section \ref{sec:core}), 5 classes are defined for CC [1,2,3,4,5] and 6 for HD [1-5,6-10,11-15,16-20,21-25,25-30], where HD are divided in groups of 5. Note that 0 or floats with 0 as the integer part are discarded for not being considered representative enough.
% halstein por debajo de 30 representa: (quitando outliers)
%91.73534499510995 diverse
%93.10333774947647 prime
%97.28381868376525 Big-Vull

%quitando outliers en cyclomatic:
%84.85344763676311 diverse
%86.03095356964553 prime
%91.95396369301562 Big-Vull

\begin{figure}
\centering
\begin{subfigure}[b]{0.49\textwidth} % Set the width to 45% of the line width for each subfigure
    \centering % Center the image within the subfigure
    \includegraphics[width=\textwidth]{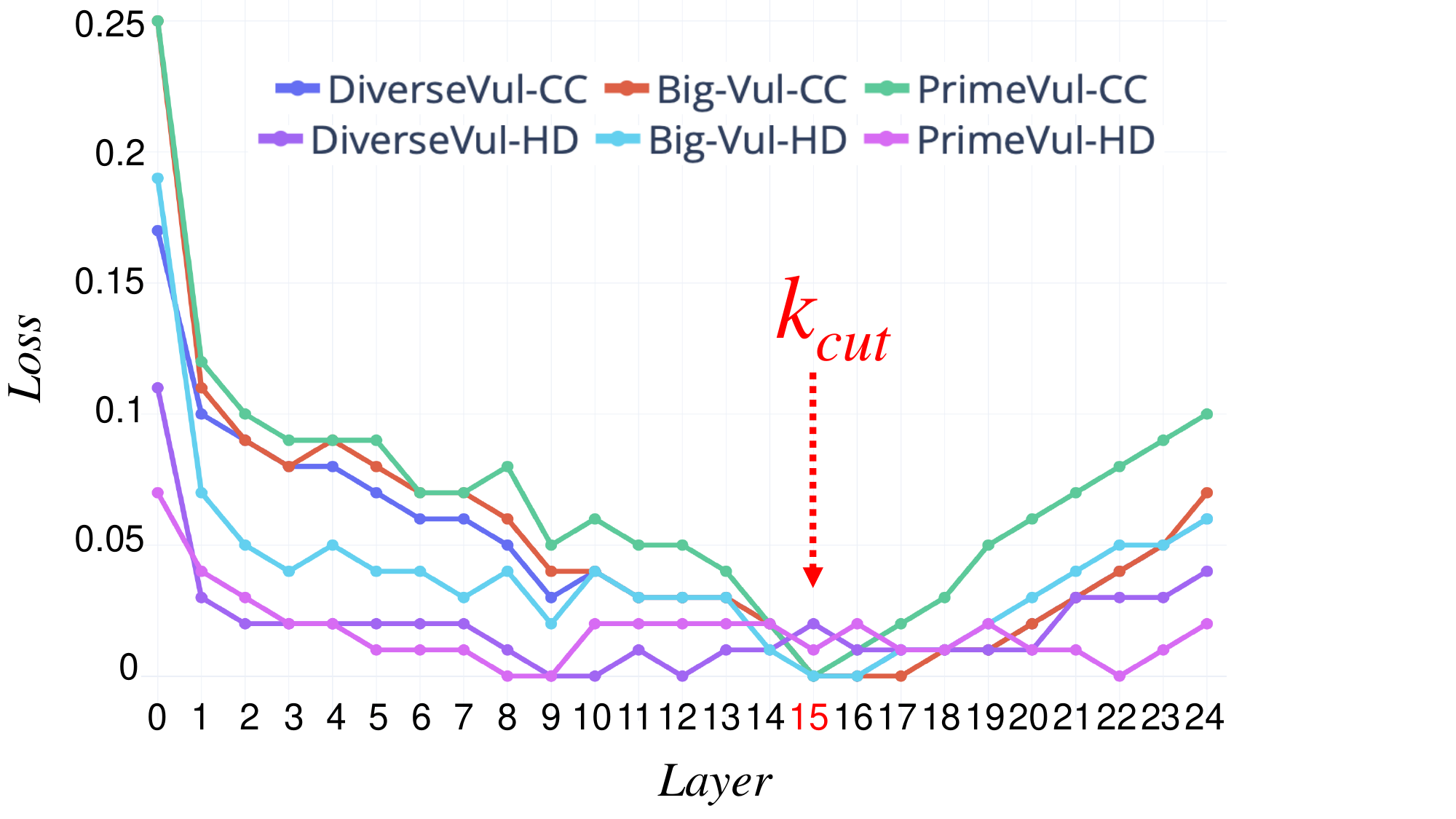} % Use \textwidth to make the image fit the subfigure width
    \caption{BERT}
    \label{fig:BertLoss}
\end{subfigure}
\hfill
\begin{subfigure}[b]{0.49\textwidth} % Use the same width for the second subfigure
    \centering % Center the image within the subfigure
    \includegraphics[width=\textwidth]{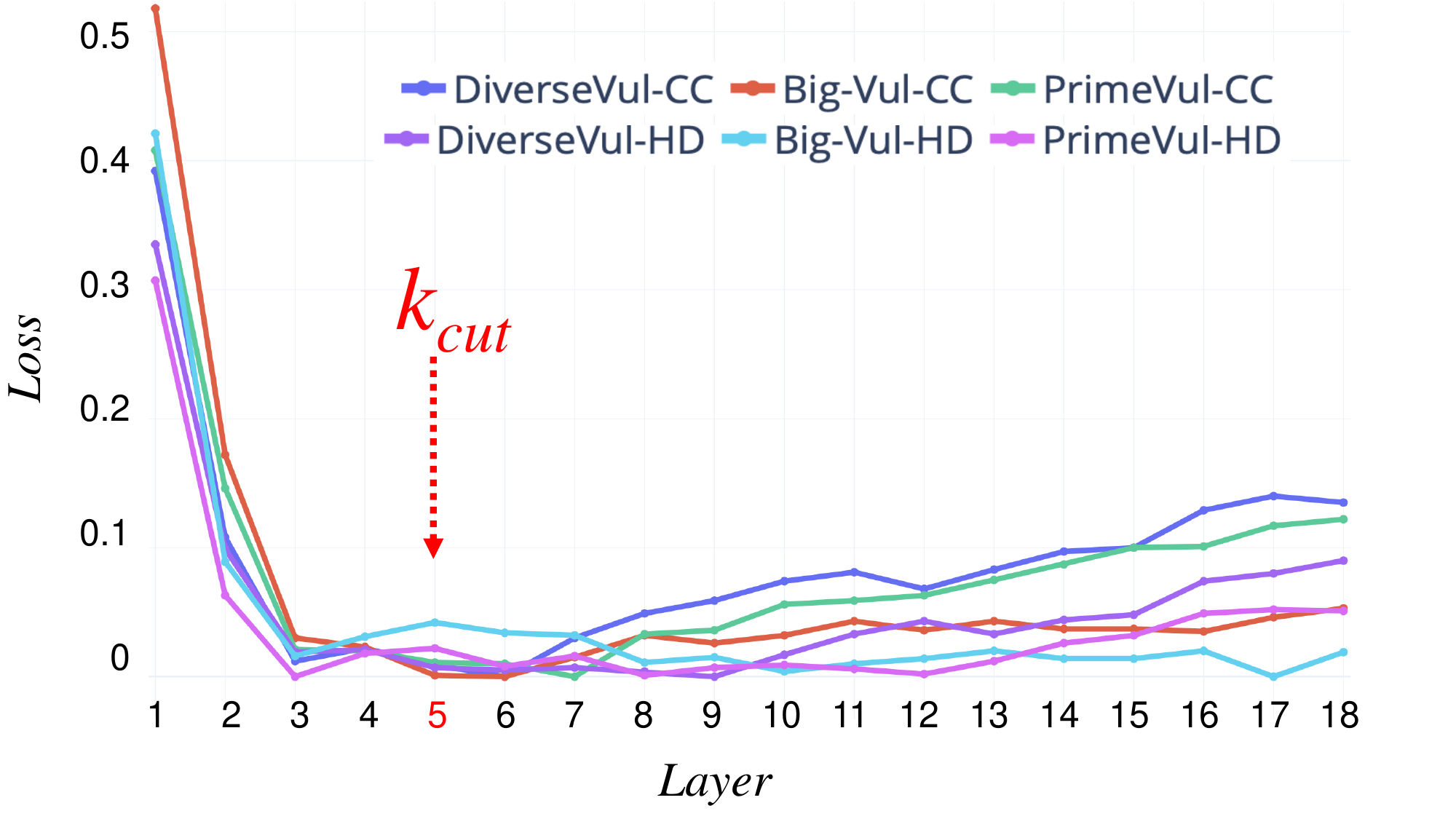} % Use \textwidth to make the image fit the subfigure width
    \caption{Gemma}
    \label{fig:GemmaLoss}
\end{subfigure}
        
\caption{LLMs $loss$ in absolute value for $D$, CC, and HD. Notice the decrement until the $K_{cut}$ layer and the later increment after that layer denoting that the hidden representations are not adding extra information to the embeddings.}
\label{fig:lossLLMsLPlayers}
\end{figure}

Once having classes, LLMs are applied and LPs are computed for all datasets $D$ and both features $F$ (i.e., CC and HD). For the sake of fairness, the same number of samples per CWE class is considered -- it is set considering the class with the minimum number of samples and downsampling the remaining classes. Figure \ref{fig:lossLLMsLPlayers} presents the $loss$ in all cases. In the case of BERT (Figure \ref{fig:lossLLMsLPlayers}-A), it is identified that, specially in CC, $loss$ gets the minimum, in all cases, around $k$=15. In HD the trend is not so clear, but  $loss$ tends to remain constant at $k$=15 or increase, in the case of Big-Vul. Then, based on Equation \ref{kcut}, $k_{cut}$=15 including the embedding layer. On the contrary, in the case of Gemma (Figure \ref{fig:lossLLMsLPlayers}-B), $loss$ shows that from $k$=3 on there are some changes in trends, being around $k$=5 and 6 when $loss$ becomes quite constant or even decrease, reaching $k_{cut}$=5. In both models CC is the code feature which presents more differences among $k$ and shows a clear trend, probably because the complexity it involves becomes harder to predict.

\subsubsection{Vulnerability detection}
\label{VulnDetec}
This section presents vulnerability detection results of all datasets $D$, BERT and Gemma (Tables \ref{tab:resultsVulDetectPruning} and \ref{tab:resultsVulDetectQuantPruning}). Effectiveness metrics are computed for identifying CWEs and for distinguishing if a code sample is or not vulnerable. 
%In particular, Tables \ref{tab:resultsVulDetectPruning} and \ref{tab:resultsVulDetectQuantPruning} depict effectiveness metrics for original models (baseline) highlighted in gray and applied compression techniques, namely quantization of 4 and 8 bits, layer pruning and the combination of pruning and quantization. Note that except for baseline results in which values are included, the percentage of increment or decrement against the baseline one is presented in all metrics. Negative values in effectiveness metrics show best results meaning that they improve after compression techniques are applied.

BERT model shows no meaningful difference compared to the baseline values when discerning across different datasets, CWEs or discerning among vulnerable and no vulnerable samples. Detecting CWE and considering just layer pruning, just the recall in DiverseVul and Big-Vul degrades more than 1\%. When applying quantization, while the difference is minimum, the pruned model shows more degradation than the non-pruned counterpart. For instance, DiverseVul and quantization 8 bits yields a degradation of 0.32\% and the pruned one 0.59\%. Similarly, in case of distinguishing vulnerable and no vulnerable samples results are almost equivalent to baseline ones with a degradation of F1 of less than 1\%.

For Gemma model results are even more successfully, as there are more negative values, which shows improvement concerning baseline results. In some cases such as in DiverseVul and Big-Vul when using a rank of 1,024, the pruned model shows better performance than the baseline. When applying quantization, with 4 bits is the only configuration that should be discarded if layer pruning is not applied, as the degradation is significant, for instance, 22.38\% of accuracy in PrimeVul. On the other hand, when distinguishing among vulnerable and no vulnerable samples, excluding the case of Big-Vul with a rank of 256, every other set-up and dataset shows a light improvement with respect to the baseline.

\paragraph{Analysis per CWE} {\color{black}Tables \ref{tab:gemma1} and \ref{tab:gemma2}  show} two confusion matrices of Gemma, for the sake of illustration (the remaining ones are available in our repository) for the {\color{black}DiverseVul and BigVul datasets}. Among the common CWEs, CWE-119 and CWE-20 generally show worse results across the datasets and models, with precision usually below the mean of the other CWEs with precisions of 76\%/78\% in BERT and 76.5\%/ 74.5\% in Gemma.

\begin{table}[htbp]
  \centering
  \begin{minipage}{0.48\textwidth}
    \centering
    \caption{{\color{black}Gemma, DiverseVul, no pruning, no quantization}}
    \scalebox{0.35}{
      \begin{tabular}{|c|c|c|c|c|c|c|c|c|c|c|c|}
        \hline
    \textbf{CWE-119} & \cellcolor[rgb]{ .6,  .6,  .6}623 & \cellcolor[rgb]{ .643,  .643,  .643}72 & \cellcolor[rgb]{ .792,  .792,  .792}42 & \cellcolor[rgb]{ .776,  .776,  .776}45 & \cellcolor[rgb]{ .961,  .961,  .961}8 & \cellcolor[rgb]{ .812,  .812,  .812}38 & \cellcolor[rgb]{ .867,  .867,  .867}27 & \cellcolor[rgb]{ .812,  .812,  .812}38 & \cellcolor[rgb]{ .941,  .941,  .941}12 & \cellcolor[rgb]{ .796,  .796,  .796}41 & \cellcolor[rgb]{ .733,  .733,  .733}54 \bigstrut\\
    \hline
    \textbf{CWE-125} & \cellcolor[rgb]{ .824,  .824,  .824}36 & \cellcolor[rgb]{ .6,  .6,  .6}714 & \cellcolor[rgb]{ .918,  .918,  .918}17 & \cellcolor[rgb]{ .843,  .843,  .843}32 & \cellcolor[rgb]{ .953,  .953,  .953}10 & \cellcolor[rgb]{ .961,  .961,  .961}8 & \cellcolor[rgb]{ .745,  .745,  .745}51 & \cellcolor[rgb]{ .808,  .808,  .808}39 & \cellcolor[rgb]{ .918,  .918,  .918}17 & \cellcolor[rgb]{ .725,  .725,  .725}55 & \cellcolor[rgb]{ .898,  .898,  .898}21 \bigstrut\\
    \hline
    \textbf{CWE-190} & \cellcolor[rgb]{ .882,  .882,  .882}24 & \cellcolor[rgb]{ .847,  .847,  .847}31 & \cellcolor[rgb]{ .6,  .6,  .6}795 & \cellcolor[rgb]{ .851,  .851,  .851}30 & \cellcolor[rgb]{ .945,  .945,  .945}11 & \cellcolor[rgb]{ .906,  .906,  .906}19 & \cellcolor[rgb]{ .922,  .922,  .922}16 & \cellcolor[rgb]{ .867,  .867,  .867}27 & \cellcolor[rgb]{ .988,  .988,  .988}3 & \cellcolor[rgb]{ .878,  .878,  .878}25 & \cellcolor[rgb]{ .906,  .906,  .906}19 \bigstrut\\
    \hline
    \textbf{CWE-20} & \cellcolor[rgb]{ .8,  .8,  .8}40 & \cellcolor[rgb]{ .851,  .851,  .851}30 & \cellcolor[rgb]{ .886,  .886,  .886}23 & \cellcolor[rgb]{ .6,  .6,  .6}687 & \cellcolor[rgb]{ .914,  .914,  .914}18 & \cellcolor[rgb]{ .745,  .745,  .745}51 & \cellcolor[rgb]{ .812,  .812,  .812}38 & \cellcolor[rgb]{ .796,  .796,  .796}41 & \cellcolor[rgb]{ .976,  .976,  .976}5 & \cellcolor[rgb]{ .906,  .906,  .906}19 & \cellcolor[rgb]{ .761,  .761,  .761}48 \bigstrut\\
    \hline
    \textbf{CWE-22} & \cellcolor[rgb]{ .992,  .992,  .992}2 & \cellcolor[rgb]{ .988,  .988,  .988}3 & \cellcolor[rgb]{ .98,  .98,  .98}4 & \cellcolor[rgb]{ .941,  .941,  .941}12 & \cellcolor[rgb]{ .6,  .6,  .6}326 & \cellcolor[rgb]{ .996,  .996,  .996}1 & \cellcolor[rgb]{ .961,  .961,  .961}8 & \cellcolor[rgb]{ .992,  .992,  .992}2 & \cellcolor[rgb]{ .976,  .976,  .976}5 & \cellcolor[rgb]{ .98,  .98,  .98}4 & \cellcolor[rgb]{ .988,  .988,  .988}3 \bigstrut\\
    \hline
    \textbf{CWE-362} & \cellcolor[rgb]{ .945,  .945,  .945}11 & \cellcolor[rgb]{ .988,  .988,  .988}3 & \cellcolor[rgb]{ .922,  .922,  .922}16 & \cellcolor[rgb]{ .863,  .863,  .863}28 & \cellcolor[rgb]{ .973,  .973,  .973}6 & \cellcolor[rgb]{ .6,  .6,  .6}853 & \cellcolor[rgb]{ .816,  .816,  .816}37 & \cellcolor[rgb]{ .914,  .914,  .914}18 & \cellcolor[rgb]{ .996,  .996,  .996}1 & \cellcolor[rgb]{ .98,  .98,  .98}4 & \cellcolor[rgb]{ .886,  .886,  .886}23 \bigstrut\\
    \hline
    \textbf{CWE-416} & \cellcolor[rgb]{ .902,  .902,  .902}20 & \cellcolor[rgb]{ .816,  .816,  .816}37 & \cellcolor[rgb]{ .867,  .867,  .867}27 & \cellcolor[rgb]{ .914,  .914,  .914}18 & \cellcolor[rgb]{ .937,  .937,  .937}13 & \cellcolor[rgb]{ .745,  .745,  .745}51 & \cellcolor[rgb]{ .6,  .6,  .6}740 & \cellcolor[rgb]{ .863,  .863,  .863}28 & \cellcolor[rgb]{ .922,  .922,  .922}16 & \cellcolor[rgb]{ .898,  .898,  .898}21 & \cellcolor[rgb]{ .859,  .859,  .859}29 \bigstrut\\
    \hline
    \textbf{CWE-476} & \cellcolor[rgb]{ .827,  .827,  .827}35 & \cellcolor[rgb]{ .753,  .753,  .753}50 & \cellcolor[rgb]{ .957,  .957,  .957}9 & \cellcolor[rgb]{ .788,  .788,  .788}43 & \cellcolor[rgb]{ .941,  .941,  .941}12 & \cellcolor[rgb]{ .851,  .851,  .851}30 & \cellcolor[rgb]{ .741,  .741,  .741}52 & \cellcolor[rgb]{ .6,  .6,  .6}686 & \cellcolor[rgb]{ .961,  .961,  .961}8 & \cellcolor[rgb]{ .725,  .725,  .725}55 & \cellcolor[rgb]{ .902,  .902,  .902}20 \bigstrut\\
    \hline
    \textbf{CWE-78} & \cellcolor[rgb]{ .992,  .992,  .992}2 & \cellcolor[rgb]{ .988,  .988,  .988}3 & \cellcolor[rgb]{ .996,  .996,  .996}1 & \cellcolor[rgb]{ .941,  .941,  .941}12 & \cellcolor[rgb]{ .988,  .988,  .988}3 & 0     & \cellcolor[rgb]{ .933,  .933,  .933}14 & \cellcolor[rgb]{ .992,  .992,  .992}2 & \cellcolor[rgb]{ .6,  .6,  .6}264 & \cellcolor[rgb]{ .996,  .996,  .996}1 & \cellcolor[rgb]{ .988,  .988,  .988}3 \bigstrut\\
    \hline
    \textbf{CWE-787} & \cellcolor[rgb]{ .722,  .722,  .722}56 & \cellcolor[rgb]{ .761,  .761,  .761}48 & \cellcolor[rgb]{ .925,  .925,  .925}15 & \cellcolor[rgb]{ .8,  .8,  .8}40 & \cellcolor[rgb]{ .992,  .992,  .992}2 & \cellcolor[rgb]{ .953,  .953,  .953}10 & \cellcolor[rgb]{ .808,  .808,  .808}39 & \cellcolor[rgb]{ .8,  .8,  .8}40 & \cellcolor[rgb]{ .961,  .961,  .961}8 & \cellcolor[rgb]{ .6,  .6,  .6}718 & \cellcolor[rgb]{ .882,  .882,  .882}24 \bigstrut\\
    \hline
    \textbf{No CWE} & \cellcolor[rgb]{ .678,  .678,  .678}65 & \cellcolor[rgb]{ .753,  .753,  .753}50 & \cellcolor[rgb]{ .902,  .902,  .902}20 & \cellcolor[rgb]{ .612,  .612,  .612}78 & \cellcolor[rgb]{ .922,  .922,  .922}16 & \cellcolor[rgb]{ .725,  .725,  .725}55 & \cellcolor[rgb]{ .6,  .6,  .6}112 & \cellcolor[rgb]{ .859,  .859,  .859}29 & \cellcolor[rgb]{ .969,  .969,  .969}7 & \cellcolor[rgb]{ .886,  .886,  .886}23 & \cellcolor[rgb]{ .6,  .6,  .6}545 \bigstrut\\
    \hline
    \rowcolor[rgb]{ .8,  .8,  .8}       & \cellcolor[rgb]{ 1,  1,  1}\textbf{CWE-119} & \cellcolor[rgb]{ 1,  1,  1}\textbf{CWE-125} & \cellcolor[rgb]{ 1,  1,  1}\textbf{CWE-190} & \cellcolor[rgb]{ 1,  1,  1}\textbf{CWE-20} & \cellcolor[rgb]{ 1,  1,  1}\textbf{CWE-22} & \cellcolor[rgb]{ 1,  1,  1}\textbf{CWE-362} & \cellcolor[rgb]{ 1,  1,  1}\textbf{CWE-416} & \cellcolor[rgb]{ 1,  1,  1}\textbf{CWE-476} & \cellcolor[rgb]{ 1,  1,  1}\textbf{CWE-78} & \cellcolor[rgb]{ 1,  1,  1}\textbf{CWE-787} & \cellcolor[rgb]{ 1,  1,  1}\textbf{No CWE} \bigstrut\\
    \hline
      \end{tabular}
       \label{tab:gemma1}%
    }
  \end{minipage}%
  \hfill
  \begin{minipage}{0.48\textwidth}
    \centering
    \caption{{\color{black}Gemma, DiverseVul, with pruning and quantization}}
    \scalebox{0.35}{
      \begin{tabular}{|c|c|c|c|c|c|c|c|c|c|c|c|}
      \hline
    \multicolumn{1}{|c|}{\textbf{CWE-119}} & \cellcolor[rgb]{ .6,  .6,  .6}646 & \cellcolor[rgb]{ .808,  .808,  .808}39 & \cellcolor[rgb]{ .957,  .957,  .957}9 & \cellcolor[rgb]{ .6,  .6,  .6}148 & 0     & \cellcolor[rgb]{ .957,  .957,  .957}9 & \cellcolor[rgb]{ .682,  .682,  .682}64 & \cellcolor[rgb]{ .933,  .933,  .933}14 & \cellcolor[rgb]{ .988,  .988,  .988}3 & 0     & \cellcolor[rgb]{ .922,  .922,  .922}16 \bigstrut\\
    \hline
    \multicolumn{1}{|c|}{\textbf{CWE-125}} & \cellcolor[rgb]{ .6,  .6,  .6}99 & \cellcolor[rgb]{ .6,  .6,  .6}570 & \cellcolor[rgb]{ .957,  .957,  .957}9 & \cellcolor[rgb]{ .737,  .737,  .737}53 & 0     & \cellcolor[rgb]{ .973,  .973,  .973}6 & \cellcolor[rgb]{ .776,  .776,  .776}45 & \cellcolor[rgb]{ .937,  .937,  .937}13 & \cellcolor[rgb]{ .988,  .988,  .988}3 & 0     & \cellcolor[rgb]{ .992,  .992,  .992}2 \bigstrut\\
    \hline
    \multicolumn{1}{|c|}{\textbf{CWE-190}} & \cellcolor[rgb]{ .714,  .714,  .714}58 & \cellcolor[rgb]{ .863,  .863,  .863}28 & \cellcolor[rgb]{ .6,  .6,  .6}276 & \cellcolor[rgb]{ .69,  .69,  .69}62 & 0     & 0     & \cellcolor[rgb]{ .792,  .792,  .792}42 & \cellcolor[rgb]{ .922,  .922,  .922}16 & \cellcolor[rgb]{ .992,  .992,  .992}2 & 0     & \cellcolor[rgb]{ .992,  .992,  .992}2 \bigstrut\\
    \hline
    \multicolumn{1}{|c|}{\textbf{CWE-20}} & \cellcolor[rgb]{ .6,  .6,  .6}123 & \cellcolor[rgb]{ .937,  .937,  .937}13 & \cellcolor[rgb]{ .992,  .992,  .992}2 & \cellcolor[rgb]{ .6,  .6,  .6}678 & \cellcolor[rgb]{ .996,  .996,  .996}1 & \cellcolor[rgb]{ .906,  .906,  .906}19 & \cellcolor[rgb]{ .663,  .663,  .663}68 & \cellcolor[rgb]{ .953,  .953,  .953}10 & \cellcolor[rgb]{ .988,  .988,  .988}3 & \cellcolor[rgb]{ .988,  .988,  .988}3 & \cellcolor[rgb]{ .945,  .945,  .945}11 \bigstrut\\
    \hline
    \multicolumn{1}{|c|}{\textbf{CWE-269}} & \cellcolor[rgb]{ .969,  .969,  .969}7 & 0     & \cellcolor[rgb]{ .996,  .996,  .996}1 & \cellcolor[rgb]{ .973,  .973,  .973}6 & \cellcolor[rgb]{ .6,  .6,  .6}111 & \cellcolor[rgb]{ .996,  .996,  .996}1 & \cellcolor[rgb]{ .992,  .992,  .992}2 & 0     & \cellcolor[rgb]{ .996,  .996,  .996}1 & 0     & 0 \bigstrut\\
    \hline
    \multicolumn{1}{|c|}{\textbf{CWE-362}} & \cellcolor[rgb]{ .6,  .6,  .6}89 & \cellcolor[rgb]{ .976,  .976,  .976}5 & \cellcolor[rgb]{ .996,  .996,  .996}1 & \cellcolor[rgb]{ .6,  .6,  .6}102 & \cellcolor[rgb]{ .996,  .996,  .996}1 & \cellcolor[rgb]{ .6,  .6,  .6}444 & \cellcolor[rgb]{ .6,  .6,  .6}111 & \cellcolor[rgb]{ .906,  .906,  .906}19 & 0     & \cellcolor[rgb]{ .992,  .992,  .992}2 & \cellcolor[rgb]{ .98,  .98,  .98}4 \bigstrut\\
    \hline
    \multicolumn{1}{|c|}{\textbf{CWE-416}} & \cellcolor[rgb]{ .631,  .631,  .631}74 & \cellcolor[rgb]{ .922,  .922,  .922}16 & 0     & \cellcolor[rgb]{ .682,  .682,  .682}64 & 0     & \cellcolor[rgb]{ .898,  .898,  .898}21 & \cellcolor[rgb]{ .6,  .6,  .6}673 & \cellcolor[rgb]{ .976,  .976,  .976}5 & \cellcolor[rgb]{ .988,  .988,  .988}3 & \cellcolor[rgb]{ .996,  .996,  .996}1 & \cellcolor[rgb]{ .973,  .973,  .973}6 \bigstrut\\
    \hline
    \textbf{CWE-476} & \cellcolor[rgb]{ .631,  .631,  .631}74 & \cellcolor[rgb]{ .937,  .937,  .937}13 & \cellcolor[rgb]{ .957,  .957,  .957}9 & \cellcolor[rgb]{ .706,  .706,  .706}59 & 0     & \cellcolor[rgb]{ .996,  .996,  .996}1 & \cellcolor[rgb]{ .725,  .725,  .725}55 & \cellcolor[rgb]{ .6,  .6,  .6}367 & \cellcolor[rgb]{ .992,  .992,  .992}2 & 0     & \cellcolor[rgb]{ .98,  .98,  .98}4 \bigstrut\\
    \hline
    \textbf{CWE-79} & \cellcolor[rgb]{ .702,  .702,  .702}60 & \cellcolor[rgb]{ .945,  .945,  .945}11 & \cellcolor[rgb]{ .988,  .988,  .988}3 & \cellcolor[rgb]{ .827,  .827,  .827}35 & 0     & \cellcolor[rgb]{ .933,  .933,  .933}14 & \cellcolor[rgb]{ .851,  .851,  .851}30 & \cellcolor[rgb]{ .976,  .976,  .976}5 & \cellcolor[rgb]{ .6,  .6,  .6}216 & 0     & \cellcolor[rgb]{ .953,  .953,  .953}10 \bigstrut\\
    \hline
    \textbf{CWE-787} & \cellcolor[rgb]{ .953,  .953,  .953}10 & \cellcolor[rgb]{ .992,  .992,  .992}2 & 0     & \cellcolor[rgb]{ .851,  .851,  .851}30 & 0     & \cellcolor[rgb]{ .992,  .992,  .992}2 & \cellcolor[rgb]{ .953,  .953,  .953}10 & \cellcolor[rgb]{ .996,  .996,  .996}1 & 0     & \cellcolor[rgb]{ .647,  .647,  .647}71 & \cellcolor[rgb]{ .988,  .988,  .988}3 \bigstrut\\
    \hline
    \textbf{No CWE} & \cellcolor[rgb]{ .6,  .6,  .6}200 & \cellcolor[rgb]{ .914,  .914,  .914}18 & \cellcolor[rgb]{ .98,  .98,  .98}4 & \cellcolor[rgb]{ .6,  .6,  .6}291 & 0     & \cellcolor[rgb]{ .722,  .722,  .722}56 & \cellcolor[rgb]{ .6,  .6,  .6}102 & \cellcolor[rgb]{ .969,  .969,  .969}7 & \cellcolor[rgb]{ .953,  .953,  .953}10 & 0     & \cellcolor[rgb]{ .6,  .6,  .6}279 \bigstrut\\
    \hline
    \rowcolor[rgb]{ .8,  .8,  .8}       & \cellcolor[rgb]{ 1,  1,  1}\textbf{CWE-119} & \cellcolor[rgb]{ 1,  1,  1}\textbf{CWE-125} & \cellcolor[rgb]{ 1,  1,  1}\textbf{CWE-190} & \cellcolor[rgb]{ 1,  1,  1}\textbf{CWE-20} & \cellcolor[rgb]{ 1,  1,  1}\textbf{CWE-269} & \cellcolor[rgb]{ 1,  1,  1}\textbf{CWE-362} & \cellcolor[rgb]{ 1,  1,  1}\textbf{CWE-416} & \cellcolor[rgb]{ 1,  1,  1}\textbf{CWE-476} & \cellcolor[rgb]{ 1,  1,  1}\textbf{CWE-79} & \cellcolor[rgb]{ 1,  1,  1}\textbf{CWE-787} & \cellcolor[rgb]{ 1,  1,  1}\textbf{No CWE} \bigstrut\\
    \hline
      \end{tabular}
       \label{tab:gemma2}%
    }
  \end{minipage}
\end{table}

CWE-119 refers to setting Improper Restrictions of Operations within the Bounds of a memory buffer, indicating that the models may fail when analyzing if the operations where properly restricted as the amount of possible improper operations is large. CWE-20 refers to improper Input Validation, suggesting that the models may lack knowledge of potential runtime inputs. Additionally, these two IDs are more general and abstract, encompassing more children classes.

While the mistakes across CWEs are low, these two CWEs in Gemma and BERT tend to be mixed, being some of the instances of the CWE-119 classified as CWE-20 and viceversa. Additionally, confusion matrices also shows that most of the miss-classifications are because the model shows lower effectiveness when discerning more these two abstract classes and the negative class.

CWE-787 and CWE-125, which refer to Out-of-bounds Write and Read respectively, show average precision across models and datasets with averages of 89.2\%/86.1\% in BERT and 87\%/ 82.5\% in Gemma. While these are specific types of operations within the broader class of CWE-119, they are more fine-grained, making them easier for the models to detect. The same applies to  CWE-362 (race-condition) and to CWE-190 (integer overflow) -- they are more concrete than CWE-119 and CWE-20.
%In the confusion matrix, in example, the CWE-125 has a 'high' ratio of error with CWE-119 supporting the previous idea.

%Other CWEs as the CWE-362 and the CWE-190, shows also good results across the datasets with means of 83\%/86,3\% in BERT and 82,1\%/ 85,3\% in Gemma. The both IDs are specific classes, where the  CWE-362 refers to a "race-condition"  and the CWE-190 to an integer overflow. These two are also specific and less asbtract than the  CWE-119 and CWE-20.

%Looking at the CWEs which are not common across all the datasets, the CWE-78 shows good performance in Primevul (93\%) and DiverseVul (83\%) and for instance, this CWE does not encompass any other CWE.

%In summary, BERT and Gemma show worse results with more abstract or general classes such as CWE-119 or CWE-20, which cover a wider range of operations and validations. However, for more specific CWEs such as CWE-787, CWE-125, CWE-78 ,CWE-362 and CWE-190, the performance is better, indicating that the more specific the CWE, the easier it is for the models to detect it.

%The negative class, in diverseVul, shows lower results when compared to the rest of the classes and negative classes in the other datasets, with an accuracy of 65\%/56\% versus, 78\%/73\% (Big-Vul) and 85\%/94\% (PrimeVul) in BERT and Gemma respectively. 

\paragraph{Corroborating $k_{cut}$ appropriateness}
On the one hand, tests {\color{black} in Table  \ref{tab:randomPruning} have been run using $\lfloor k_{cut} \rfloor$/2 as cut-off value, that is 5 in Gemma and 15 in BERT. Thus, the new cut-off value} is 2 and 7 for Gemma and BERT respectively, such that 15 and 16 layers are removed in each case. All models'  metrics get worse, corroborating that the proposed approach achieves a nice reduction of the model, while keeping or improving {\color{black}the results of the original models}. 

\begin{table}[htbp]
  \centering
  \caption{{\color{black}Additional layer pruning. Notice that a positive value means a decrement with respect to the baseline.}}
   \scalebox{0.55}{
    \begin{tabular}{|l|c|r|r|r|r|r|}
\cline{3-7}    \multicolumn{1}{r}{} &       & \multicolumn{5}{p{20.33em}|}{\textbf{Effectiveness}} \\
\cline{3-7}    \multicolumn{1}{r}{} &       & \multicolumn{4}{p{16.83em}|}{\textbf{CWE}} & \multicolumn{1}{p{3.5em}|}{\textbf{Vul-NoVul}} \\
    \hline
    \multicolumn{1}{|p{4.165em}|}{\textbf{Rank}} & \multicolumn{1}{p{4.335em}|}{\textbf{Removed layers}} & \multicolumn{1}{p{4.055em}|}{\textbf{Accuracy}} & \multicolumn{1}{p{3.61em}|}{\textbf{F1}} & \multicolumn{1}{p{4.11em}|}{\textbf{Precision}} & \multicolumn{1}{p{5.055em}|}{\textbf{Recall}} & \multicolumn{1}{p{3.5em}|}{\textbf{F1}} \\
    \hline
    \multicolumn{7}{|c|}{\textbf{Gemma (18 layers)}} \\
    \hline
    \multicolumn{7}{|p{28.83em}|}{\textbf{DiverseVul}} \\
    \hline
    \rowcolor[rgb]{ .949,  .949,  .949} \multicolumn{1}{|c|}{256} & 0     & \multicolumn{1}{c|}{73.40\%} & \multicolumn{1}{c|}{74.90\%} & \multicolumn{1}{c|}{75.30\%} & \multicolumn{1}{c|}{75.00\%} & \multicolumn{1}{c|}{95.90\%} \\
    \hline
    \multicolumn{1}{|c|}{\textit{256}} & \textit{15} & \multicolumn{1}{p{4.055em}|}{\textit{2,50}} & \multicolumn{1}{p{3.61em}|}{\textit{2,26}} & \multicolumn{1}{p{4.11em}|}{\textit{2,48}} & \multicolumn{1}{p{5.055em}|}{\textit{2,29}} & \multicolumn{1}{p{3.5em}|}{\textit{0,48}} \\
    \hline
    \multicolumn{1}{|c|}{\textit{1024}} & \textit{15} & \multicolumn{1}{p{4.055em}|}{\textit{2,82}} & \multicolumn{1}{p{3.61em}|}{\textit{3,01}} & \multicolumn{1}{p{4.11em}|}{\textit{3,04}} & \multicolumn{1}{p{5.055em}|}{\textit{2,64}} & \multicolumn{1}{p{3.5em}|}{\textit{0,19}} \\
    \hline
    \multicolumn{7}{|p{28.83em}|}{\textbf{PrimeVul}} \\
    \hline
    \rowcolor[rgb]{ .949,  .949,  .949} \multicolumn{1}{|c|}{256} & 0     & \multicolumn{1}{c|}{88.10\%} & \multicolumn{1}{c|}{89.00\%} & \multicolumn{1}{c|}{88.80\%} & \multicolumn{1}{c|}{88.10\%} & \multicolumn{1}{c|}{99.50\%} \\
    \hline
    \multicolumn{1}{|c|}{\textit{256}} & \textit{15} & \multicolumn{1}{p{4.055em}|}{\textit{3,00}} & \multicolumn{1}{p{3.61em}|}{\textit{2,10}} & \multicolumn{1}{p{4.11em}|}{\textit{0,80}} & \multicolumn{1}{p{5.055em}|}{\textit{2,11}} & \multicolumn{1}{p{3.5em}|}{\textit{0,28}} \\
    \hline
    \multicolumn{1}{|c|}{\textit{1024}} & \textit{15} & \multicolumn{1}{p{4.055em}|}{\textit{4,20}} & \multicolumn{1}{p{3.61em}|}{\textit{3,63}} & \multicolumn{1}{p{4.11em}|}{\textit{3,23}} & \multicolumn{1}{p{5.055em}|}{\textit{2,60}} & \multicolumn{1}{p{3.5em}|}{\textit{0,17}} \\
    \hline
    \multicolumn{7}{|p{28.83em}|}{\textbf{Big-Vul}} \\
    \hline
    \rowcolor[rgb]{ .949,  .949,  .949} \multicolumn{1}{|c|}{256} & 0     & \multicolumn{1}{c|}{77.00\%} & \multicolumn{1}{c|}{79.30\%} & \multicolumn{1}{c|}{78.80\%} & \multicolumn{1}{c|}{79.90\%} & \multicolumn{1}{c|}{95.30\%} \\
    \hline
    \multicolumn{1}{|c|}{\textit{256}} & \textit{15} & \multicolumn{1}{p{4.055em}|}{\textit{2,46}} & \multicolumn{1}{p{3.61em}|}{\textit{1,37}} & \multicolumn{1}{p{4.11em}|}{\textit{0,49}} & \multicolumn{1}{p{5.055em}|}{\textit{2,01}} & \multicolumn{1}{p{3.5em}|}{\textit{1,48}} \\
    \hline
    \multicolumn{1}{|c|}{\textit{1024}} & \textit{15} & \multicolumn{1}{p{4.055em}|}{\textit{6,96}} & \multicolumn{1}{p{3.61em}|}{\textit{5,28}} & \multicolumn{1}{p{4.11em}|}{\textit{5,46}} & \multicolumn{1}{p{5.055em}|}{\textit{4,53}} & \multicolumn{1}{p{3.5em}|}{\textit{2,18}} \\
    \hline
    \multicolumn{7}{|c|}{\textbf{BERT (24 layers)}} \\
    \hline
    \multicolumn{7}{|p{28.83em}|}{\textbf{DiverseVul}} \\
    \hline
    \rowcolor[rgb]{ .949,  .949,  .949} -     & 0     & 78.20\% & 79.60\% & 79.50\% & 79.60\% & 96.10\% \\
    \hline
    -     & \textit{16} & \textit{4.1} & \textit{3.73} & \textit{4.5} & \textit{3.27} & \textit{1.02} \\
    \hline
    \multicolumn{7}{|p{28.83em}|}{\textbf{PrimeVul}} \\
    \hline
    \rowcolor[rgb]{ .949,  .949,  .949} -     & 0     & 86.10\% & 86.90\% & 87.10\% & 86.60\% & 99.20\% \\
    \hline
    -     & \textit{16} & \textit{2} & \textit{1.9} & \textit{1.81} & \textit{1.8} & \textit{2.99} \\
    \hline
    \multicolumn{7}{|p{28.83em}|}{\textbf{Big-Vul}} \\
    \hline
    \rowcolor[rgb]{ .949,  .949,  .949} -     & 0     & 82.40\% & 85.10\% & 85.40\% & 84.70\% & 95.90\% \\
    \hline
    -     & \textit{16} & \textit{1.44} & \textit{1.05} & \textit{1.43} & \textit{0.28} & \textit{0.55} \\
    \hline
    \end{tabular}%
	}
  \label{tab:randomPruning}%
\end{table}%

\begin{figure}[hbtp!]
    \centering
   \includegraphics[width=0.4\textwidth]{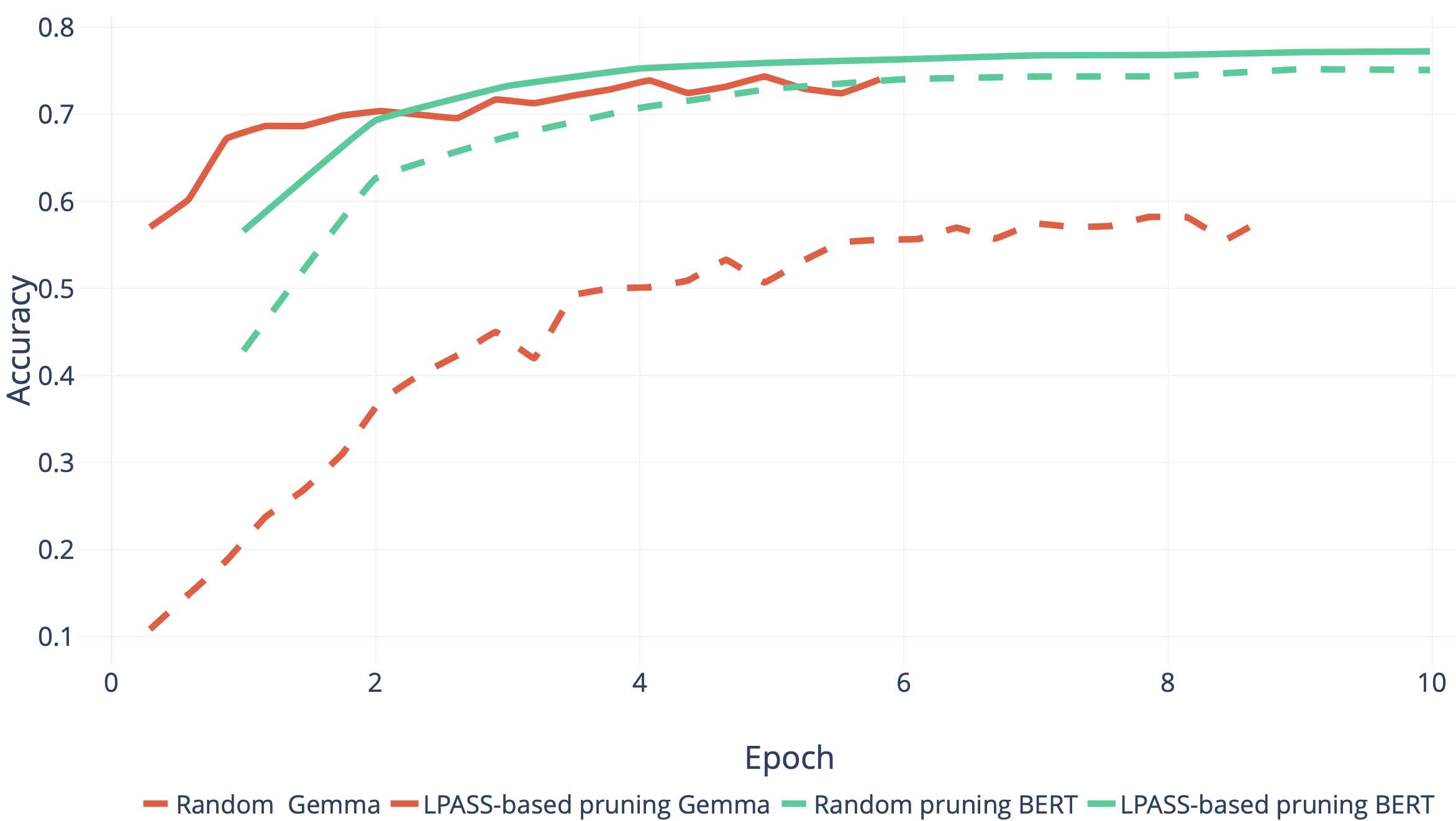}
    \caption{{\color{black}Comparison of random pruning vs $LPASS$-based pruning. Dotted line represents the random pruning version of their respective models }}
    \label{fig:randomVSnormal}
\end{figure}

On the other hand, a comparison against random layer removal (as done in \cite{sajjad2023effect}) has been carried out in {\color{black} Figure \ref{fig:randomVSnormal}}. $k_{cut}$ random layers are removed, and repeated 3 times, from BERT and Gemma to analyse average results afterwards. Specially in Gemma results are quite worse removing random layers than using our proposed $k_{cut}$ with a reduction of 20\% of accuracy in the best case. BERT, though  to a lesser degree, also presents a degradation of accuracy when applying random pruning, that is around 10\% and 2.5\% in the worst and best case respectively.

%Figure \ref{fig:randomVSnormal} depicts the accuracy evolution of Gemma and BERT after being pruned with random and LP guided pruning method. Gemma's accuracy exhibits a pronounced difference between the two pruning techniques, The dotted line represents the results of random pruning. Additionally, BERT's accuracy evolution, depicted in green, also shows degradation between the linear probe and the random  pruning method counterparts. However, the degradation is lower. This points the importance of the layers sequentially when pruning before fine-tuning.

%Therefore, LP guiding method helps on preserving better the representation of the code across the layers obtained during pre-training.

\subsubsection{Time and memory efficiency}
\label{TMefficiency}
Tables \ref{tab:resultsVulDetectPruning} and \ref{tab:resultsVulDetectQuantPruning} also show these performance metrics. %also depict performance metrics with baseline results highlighted in gray, though in this case the interpretation is the opposite as in vulnerability detection. As these metrics should be minimized, positive values are preferred as they represent a decrease. 
For BERT, the memory required for full fine-tuning on a consumer-grade GPU is reduced by 27.7\%, making it more accessible for lower-budget GPUs. The memory required for inference also decreases by around 15.41\%, while  35.7\%. parameters and 26.32\% the model size due to the need to store fewer weights. This also translates to better inference time, with an improvement across all datasets of 32\%.

When BERT is quantized, inference time is slowed down, especially on 8-bit versions. Nonetheless, pruning alleviates this problem, reducing the slowdown from -846\% (non-pruned DiverseVul) to -517\% (pruned 8-bit DiverseVul). This could be due to the extra overhead and operations required by the 8-bit and 4-bit linear layers. Despite this increase in inference time, BERT model benefits from quantization by reducing the model size and the effective parameters further. A 4-bit pruned BERT reduces the original model size by 86.73\% with a reduction of 164\% in effective parameters. In terms of GPU memory, the requirements are significantly reduced, especially in the 8-bit versions, which, after pruning, are reduced by 30\% across the datasets.

Gemma also shows a positive impact on its metrics thanks to pruning, reducing the training time by 71.36\% in PrimeVul with a rank of 1,024. These gains are also seen in inference time, with reductions of 78\%. While higher ranks require more GPU memory for fine-tuning, a rank of 1024 reduces memory usage by 34.98\%, and a rank of 256 41\%. Pruning the model also benefits memory usage for inference, which drops by 49\%, primarily due to a 52.72\% reduction in effective parameters. This results in model sizes being 57\% smaller.

As with in previous model, quantizing the Gemma model translates to worse inference times. However, pruned Gemma alleviates this issue due to the smaller model complexity, reducing the increase from -2130.77\% (8-bit PrimeVul non-pruned) to -223.08\% (8-bit PrimeVul pruned). %{\color{red}XXLUIS: Add something here. -2000\% is WAY TOO MUCH just to say "well, we are slower"}.
Additionally, quantized Gemma reduces the original model size by 72\% in the 4-bit version, which can be explained by the reduction of effective parameters by around 68\% across the datasets.

Overall, performance metrics show that training time is reduced by almost 30\% in PrimeVul and Big-Vul and more than 50\% in DiverseVul, while inference time decreases by 28\% across all datasets and models. Model metrics also improve similarly across all datasets, with around a 30\% improvement in each. %{\color{black}While quantization increases the overhead at inference with the quantized linear layers, it allows to compress and reduce the model size and the number of effective parameters even further while keeping accuracy.
%}

%{\color{red}Luis: add here values of LPs in terms of time. Compute the saving (i.e., time of computing LPs - time of fine-tuning). Stress the value of this gain}

\subsubsection{Early assessment}
\label{EarlyAssess}
The performance estimation process requires computing $\mathcal{E}(D)$. First, Table \ref{tab:estimateBeta} presents values of $\beta$ for combinations of $\mathcal{D}$ applying the leave-one-out cross-validation -- a couple of datasets are used for computing $\beta$, and the remaining one to compute $\mathcal{E}(D)$. Interestingly, it can be seen that $\beta$ is reasonably stable for different datasets, which is beneficial for supporting the generalization of this approach.

Recalling Equation \ref{EE}, the combination of $\beta$ and $\mathcal{E}(D)$ allows computing $Err$ in each case. Table \ref{tab:estimate2} depicts $Err$ for all $D$, original models and compression configurations, namely original models, models with 4 and 8 bits quantization, after established layer pruning and pruning together with quantization.

BERT produces the best results with an $Err$ between 8\% and 4\% and an average of 4.6\% in CC and 7\% in HD. On the other hand, Gemma presents a bit worse results, specially for CC in which the $Err$ is between 17\% and 11\% and an average of 13.1\%, while in CC the average $Err$ in 10.2\%. Moreover, there is not a clear distinction of LLM configurations as $Err$ remains similar among them.

\begin{table*}[htbp]
  \centering
  \caption{Values for $\beta$ (in \%) for each LLM setup phase }
  \scalebox{0.5}{
    \begin{tabular}{|p{5.39em}|r|r|r|r|r|r|r|r|r|r|r|r|r|r|r|r|r|r|}
\cline{2-19}    \multicolumn{1}{r|}{} & \multicolumn{3}{p{9.725em}|}{\textbf{Baseline}} & \multicolumn{3}{p{9.335em}|}{\textcolor[rgb]{ .122,  .122,  .122}{\textbf{Quantization 4 bits}}} & \multicolumn{3}{p{9.725em}|}{\textcolor[rgb]{ .122,  .122,  .122}{\textbf{Quantization 8 bits}}} & \multicolumn{3}{p{9.78em}|}{\textbf{Layer pruning}} & \multicolumn{3}{p{8.775em}|}{\textcolor[rgb]{ .122,  .122,  .122}{\textbf{L. Pruning+ Quant 4 bits}}} & \multicolumn{3}{p{8.775em}|}{\textcolor[rgb]{ .122,  .122,  .122}{\textbf{L. Pruning+ Quant 8 bits}}} \\
\cline{2-19}    \multicolumn{1}{r|}{} & \multicolumn{1}{p{4.39em}|}{\textbf{Precision}} & \multicolumn{1}{p{3em}|}{\textbf{Recall}} & \multicolumn{1}{p{2.335em}|}{\textbf{F1}} & \multicolumn{1}{p{4.445em}|}{\textbf{Precision}} & \multicolumn{1}{p{3em}|}{\textbf{Recall}} & \multicolumn{1}{p{1.89em}|}{\textbf{F1}} & \multicolumn{1}{p{4.5em}|}{\textbf{Precision}} & \multicolumn{1}{p{2.945em}|}{\textbf{Recall}} & \multicolumn{1}{p{2.28em}|}{\textbf{F1}} & \multicolumn{1}{p{4.335em}|}{\textbf{Precision}} & \multicolumn{1}{p{3em}|}{\textbf{Recall}} & \multicolumn{1}{p{2.445em}|}{\textbf{F1}} & \multicolumn{1}{p{4.165em}|}{\textbf{Precision}} & \multicolumn{1}{p{2.72em}|}{\textbf{Recall}} & \multicolumn{1}{p{1.89em}|}{\textbf{F1}} & \multicolumn{1}{p{4.165em}|}{\textbf{Precision}} & \multicolumn{1}{p{2.72em}|}{\textbf{Recall}} & \multicolumn{1}{p{1.89em}|}{\textbf{F1}} \\
    \hline
    \multicolumn{19}{|c|}{\textbf{BERT}} \\
    \hline
    \multicolumn{19}{|c|}{\textbf{CC}} \\
    \hline
    DiverseVul+ PrimeVul & 38    & 38    & 38    & 38    & 38    & 38    & 38    & 38    & 38    & 38    & 38    & 38    & 37    & 37    & 37    & 38    & 38    & 38 \\
    \hline
    PrimeVul+ Big-Vul & 38    & 38    & 38    & 38    & 38    & 38    & 38    & 38    & 38    & 38    & 38    & 38    & 37    & 37    & 37    & 38    & 38    & 38 \\
    \hline
    DiverseVul+ Big-Vul & 37    & 37    & 37    & 36    & 37    & 37    & 37    & 37    & 37    & 36    & 37    & 36    & 35    & 36    & 35    & 36    & 36    & 36 \\
    \hline
    \textbf{All D} & 38    & 38    & 38    & 37    & 37    & 37    & 38    & 38    & 38    & 37    & 37    & 37    & 37    & 37    & 36    & 37    & 37    & 37 \\
    \hline
    \multicolumn{1}{|r|}{} & \multicolumn{18}{c|}{\textbf{HD}} \\
    \hline
    DiverseVul+ PrimeVul & 41    & 41    & 41    & 41    & 41    & 41    & 41    & 41    & 41    & 40    & 40    & 40    & 39    & 39    & 39    & 40    & 40    & 40 \\
    \hline
    PrimeVul+ Big-Vul & 41    & 40    & 40    & 37    & 40    & 40    & 41    & 40    & 41    & 40    & 40    & 40    & 40    & 40    & 40    & 40    & 40    & 40 \\
    \hline
    DiverseVul+ Big-Vul & 37    & 37    & 37    & 33    & 36    & 36    & 37    & 37    & 37    & 35    & 36    & 35    & 34    & 35    & 34    & 35    & 35    & 35 \\
    \hline
    \textbf{All D} & 40    & 40    & 39    & 37    & 39    & 39    & 40    & 39    & 40    & 38    & 39    & 38    & 38    & 38    & 38    & 38    & 38    & 38 \\
    \hline
    \hline
    \multicolumn{19}{|c|}{\textbf{Gemma}} \\
    \hline
    \multicolumn{19}{|c|}{\textbf{CC}} \\
    \hline
    DiverseVul+ PrimeVul & 28    & 27    & 27    & 10    & 4     & 6     & 28    & 28    & 28    & 24    & 27    & 28    & 25    & 19    & 21    & 28    & 27    & 27 \\
    \hline
    PrimeVul+ Big-Vul & 24    & 24    & 24    & 13    & 2     & 2     & 25    & 24    & 24    & 28    & 26    & 27    & 20    & 10    & 12    & 26    & 26    & 26 \\
    \hline
    DiverseVul+ Big-Vul & 18    & 18    & 18    & 4     & 4     & 4     & 18    & 17    & 17    & 16    & 19    & 19    & 16    & 10    & 11    & 19    & 19    & 19 \\
    \hline
    \textbf{All D} & 23    & 23    & 23    & 9     & 3     & 4     & 24    & 23    & 23    & 22    & 24    & 25    & 20    & 13    & 15    & 24    & 24    & 24 \\
    \hline
    \multicolumn{1}{|r|}{} & \multicolumn{18}{c|}{\textbf{HD}} \\
    \hline
    DiverseVul+ PrimeVul & 35    & 34    & 34    & 17    & 6     & 8     & 35    & 34    & 35    & 32    & 36    & 37    & 34    & 28    & 30    & 37    & 36    & 36 \\
    \hline
    PrimeVul+ Big-Vul & 35    & 34    & 35    & 18    & 11    & 13    & 35    & 35    & 35    & 39    & 38    & 38    & 32    & 21    & 24    & 37    & 37    & 37 \\
    \hline
    DiverseVul+ Big-Vul & 27    & 26    & 26    & 7     & 6     & 8     & 26    & 26    & 26    & 25    & 29    & 29    & 25    & 20    & 21    & 28    & 29    & 28 \\
    \hline
    \textbf{All D} & 32    & 31    & 32    & 14    & 8     & 10    & 32    & 32    & 32    & 32    & 34    & 35    & 30    & 23    & 25    & 34    & 34    & 34 \\
    \hline
    \end{tabular}%
    }
  \label{tab:estimateBeta}%
\end{table*}%

\begin{table*}[htbp]
  \centering
  \caption{$Err$ (in \%) per code feature $F$ for each LLM setup phase. Datasets are used following a leave-one-out cross-validation. Only common CWEs are considered}
   \scalebox{0.5}{
    \begin{tabular}{|c|r|p{5.39em}|r|r|r|r|r|r|r|r|r|r|r|r|r|r|r|r|r|r|}
\cline{4-21}    \multicolumn{1}{r}{} & \multicolumn{1}{r}{} & \multicolumn{1}{r|}{} & \multicolumn{3}{p{10.33em}|}{\textbf{Baseline}} & \multicolumn{3}{p{9em}|}{\textbf{Quantization 4 bits}} & \multicolumn{3}{p{10.22em}|}{\textbf{Quantization 8 bits}} & \multicolumn{3}{p{9.555em}|}{\textbf{Layer pruning}} & \multicolumn{3}{p{10.22em}|}{\textbf{L. Pruning+ Quant 4 bits}} & \multicolumn{3}{p{10.555em}|}{\textbf{L. Prun8ng+ Quant 8 bits}} \\
\cline{2-21}    \multicolumn{1}{r|}{} & \multicolumn{1}{p{5.28em}|}{\textbf{LP datasets}} & \textbf{Prediction datasets} & \multicolumn{1}{p{4.11em}|}{\textbf{Precision}} & \multicolumn{1}{p{3.555em}|}{\textbf{Recall}} & \multicolumn{1}{p{2.665em}|}{\textbf{F1}} & \multicolumn{1}{p{3.555em}|}{\textbf{Precision}} & \multicolumn{1}{p{3.555em}|}{\textbf{Recall}} & \multicolumn{1}{p{1.89em}|}{\textbf{F1}} & \multicolumn{1}{p{4.11em}|}{\textbf{Precision}} & \multicolumn{1}{p{3.555em}|}{\textbf{Recall}} & \multicolumn{1}{p{2.555em}|}{\textbf{F1}} & \multicolumn{1}{p{4.11em}|}{\textbf{Precision}} & \multicolumn{1}{p{3.555em}|}{\textbf{Recall}} & \multicolumn{1}{p{1.89em}|}{\textbf{F1}} & \multicolumn{1}{p{4.11em}|}{\textbf{Precision}} & \multicolumn{1}{p{3.555em}|}{\textbf{Recall}} & \multicolumn{1}{p{2.555em}|}{\textbf{F1}} & \multicolumn{1}{p{4.11em}|}{\textbf{Precision}} & \multicolumn{1}{p{3.555em}|}{\textbf{Recall}} & \multicolumn{1}{p{2.89em}|}{\textbf{F1}} \\
\hline  \multicolumn{21}{|c|}{\textbf{BERT}} \\
    \hline
    \multicolumn{1}{|c|}{\multirow{4}[8]{*}{CC}} & \multicolumn{1}{p{5.28em}|}{DiverseVul+ PrimeVul} & Big-Vul & 5     & 5     & 5     & 5     & 5     & 5     & 5     & 5     & 5     & 5     & 6     & 5     & 5     & 6     & 5     & 4     & 6     & 5 \\
\cline{2-21}          & \multicolumn{1}{p{5.28em}|}{PrimeVul+ Big-Vul} & DiverseVul & 5     & 5     & 5     & 5     & 5     & 5     & 5     & 5     & 5     & 5     & 6     & 6     & 6     & 7     & 6     & 6     & 6     & 6 \\
\cline{2-21}          & \multicolumn{1}{p{5.28em}|}{DiverseVul+ Big-Vul} & PrimeVul & 3     & 3     & 3     & 4     & 3     & 3     & 3     & 3     & 3     & 4     & 3     & 3     & 5     & 4     & 4     & 4     & 4     & 4 \\
\cline{2-21}          &       & \cellcolor[rgb]{ .949,  .949,  .949}\textbf{Mean} & \cellcolor[rgb]{ .949,  .949,  .949}5 & \cellcolor[rgb]{ .949,  .949,  .949}4 & \cellcolor[rgb]{ .949,  .949,  .949}4 & \cellcolor[rgb]{ .949,  .949,  .949}5 & \cellcolor[rgb]{ .949,  .949,  .949}4 & \cellcolor[rgb]{ .949,  .949,  .949}4 & \cellcolor[rgb]{ .949,  .949,  .949}4 & \cellcolor[rgb]{ .949,  .949,  .949}4 & \cellcolor[rgb]{ .949,  .949,  .949}4 & \cellcolor[rgb]{ .949,  .949,  .949}5 & \cellcolor[rgb]{ .949,  .949,  .949}5 & \cellcolor[rgb]{ .949,  .949,  .949}5 & \cellcolor[rgb]{ .949,  .949,  .949}5 & \cellcolor[rgb]{ .949,  .949,  .949}6 & \cellcolor[rgb]{ .949,  .949,  .949}5 & \cellcolor[rgb]{ .949,  .949,  .949}5 & \cellcolor[rgb]{ .949,  .949,  .949}5 & \cellcolor[rgb]{ .949,  .949,  .949}5 \\
    \hline
    \multicolumn{1}{|c|}{\multirow{4}[8]{*}{HD}} & \multicolumn{1}{p{5.28em}|}{DiverseVul+ PrimeVul} & Big-Vul & 6     & 5     & 5     & 6     & 6     & 5     & 6     & 5     & 5     & 4     & 6     & 5     & 5     & 5     & 5     & 5     & 5     & 5 \\
\cline{2-21}          & \multicolumn{1}{p{5.28em}|}{PrimeVul+ Big-Vul} & DiverseVul & 6     & 6     & 6     & 5     & 6     & 6     & 6     & 6     & 6     & 8     & 9     & 8     & 8     & 9     & 8     & 8     & 9     & 8 \\
\cline{2-21}          & \multicolumn{1}{p{5.28em}|}{DiverseVul+ Big-Vul} & PrimeVul & 8     & 7     & 8     & 12    & 8     & 8     & 8     & 7     & 8     & 10    & 9     & 9     & 10    & 10    & 10    & 9     & 9     & 9 \\
\cline{2-21}          &       & \cellcolor[rgb]{ .949,  .949,  .949}\textbf{Mean} & \cellcolor[rgb]{ .949,  .949,  .949}7 & \cellcolor[rgb]{ .949,  .949,  .949}6 & \cellcolor[rgb]{ .949,  .949,  .949}6 & \cellcolor[rgb]{ .949,  .949,  .949}8 & \cellcolor[rgb]{ .949,  .949,  .949}7 & \cellcolor[rgb]{ .949,  .949,  .949}7 & \cellcolor[rgb]{ .949,  .949,  .949}7 & \cellcolor[rgb]{ .949,  .949,  .949}6 & \cellcolor[rgb]{ .949,  .949,  .949}6 & \cellcolor[rgb]{ .949,  .949,  .949}7 & \cellcolor[rgb]{ .949,  .949,  .949}8 & \cellcolor[rgb]{ .949,  .949,  .949}7 & \cellcolor[rgb]{ .949,  .949,  .949}8 & \cellcolor[rgb]{ .949,  .949,  .949}8 & \cellcolor[rgb]{ .949,  .949,  .949}8 & \cellcolor[rgb]{ .949,  .949,  .949}7 & \cellcolor[rgb]{ .949,  .949,  .949}7 & \cellcolor[rgb]{ .949,  .949,  .949}7 \\
    \hline
    \hline
    \multicolumn{21}{|c|}{\textbf{Gemma}} \\
    \hline
    \multicolumn{1}{|c|}{\multirow{4}[8]{*}{CC}} & \multicolumn{1}{p{5.28em}|}{DiverseVul+ PrimeVul} & Big-Vul & 14    & 13    & 13    & 15    & 16    & 11    & 14    & 14    & 14    & 11    & 10    & 10    & 17    & 19    & 18    & 12    & 11    & 11 \\
\cline{2-21}          & \multicolumn{1}{p{5.28em}|}{PrimeVul+ Big-Vul} & DiverseVul & 5     & 9     & 6     & 13    & 16    & 11    & 7     & 7     & 7     & 16    & 9     & 9     & 6     & 13    & 9     & 8     & 9     & 8 \\
\cline{2-21}          & \multicolumn{1}{p{5.28em}|}{DiverseVul+ Big-Vul} & PrimeVul & 16    & 15    & 16    & 17    & 19    & 11    & 17    & 17    & 17    & 20    & 16    & 16    & 17    & 14    & 12    & 15    & 14    & 15 \\
\cline{2-21}          &       & \cellcolor[rgb]{ .949,  .949,  .949}\textbf{Mean} & \cellcolor[rgb]{ .949,  .949,  .949}11 & \cellcolor[rgb]{ .949,  .949,  .949}13 & \cellcolor[rgb]{ .949,  .949,  .949}12 & \cellcolor[rgb]{ .949,  .949,  .949}15 & \cellcolor[rgb]{ .949,  .949,  .949}17 & \cellcolor[rgb]{ .949,  .949,  .949}11 & \cellcolor[rgb]{ .949,  .949,  .949}13 & \cellcolor[rgb]{ .949,  .949,  .949}13 & \cellcolor[rgb]{ .949,  .949,  .949}13 & \cellcolor[rgb]{ .949,  .949,  .949}16 & \cellcolor[rgb]{ .949,  .949,  .949}12 & \cellcolor[rgb]{ .949,  .949,  .949}12 & \cellcolor[rgb]{ .949,  .949,  .949}14 & \cellcolor[rgb]{ .949,  .949,  .949}15 & \cellcolor[rgb]{ .949,  .949,  .949}13 & \cellcolor[rgb]{ .949,  .949,  .949}12 & \cellcolor[rgb]{ .949,  .949,  .949}12 & \cellcolor[rgb]{ .949,  .949,  .949}12 \\
    \hline
    \multicolumn{1}{|c|}{\multirow{4}[8]{*}{HD}} & \multicolumn{1}{p{5.28em}|}{DiverseVul+ PrimeVul} & Big-Vul & 8     & 6     & 6     & 10    & 10    & 6     & 7     & 6     & 7     & 5     & 6     & 5     & 12    & 13    & 12    & 6     & 7     & 6 \\
\cline{2-21}          & \multicolumn{1}{p{5.28em}|}{PrimeVul+ Big-Vul} & DiverseVul & 7     & 9     & 8     & 11    & 14    & 13    & 8     & 8     & 8     & 16    & 9     & 9     & 5     & 7     & 5     & 8     & 9     & 8 \\
\cline{2-21}          & \multicolumn{1}{p{5.28em}|}{DiverseVul+ Big-Vul} & PrimeVul & 13    & 13    & 13    & 17    & 16    & 9     & 15    & 14    & 15    & 17    & 13    & 13    & 17    & 13    & 12    & 13    & 12    & 13 \\
\cline{2-21}          &       & \cellcolor[rgb]{ .949,  .949,  .949}Mean & \cellcolor[rgb]{ .949,  .949,  .949}9 & \cellcolor[rgb]{ .949,  .949,  .949}10 & \cellcolor[rgb]{ .949,  .949,  .949}9 & \cellcolor[rgb]{ .949,  .949,  .949}13 & \cellcolor[rgb]{ .949,  .949,  .949}13 & \cellcolor[rgb]{ .949,  .949,  .949}9 & \cellcolor[rgb]{ .949,  .949,  .949}10 & \cellcolor[rgb]{ .949,  .949,  .949}10 & \cellcolor[rgb]{ .949,  .949,  .949}10 & \cellcolor[rgb]{ .949,  .949,  .949}13 & \cellcolor[rgb]{ .949,  .949,  .949}9 & \cellcolor[rgb]{ .949,  .949,  .949}9 & \cellcolor[rgb]{ .949,  .949,  .949}11 & \cellcolor[rgb]{ .949,  .949,  .949}11 & \cellcolor[rgb]{ .949,  .949,  .949}10 & \cellcolor[rgb]{ .949,  .949,  .949}9 & \cellcolor[rgb]{ .949,  .949,  .949}9 & \cellcolor[rgb]{ .949,  .949,  .949}9 \\
    \hline
    \end{tabular}%
	}
  \label{tab:estimate2}%
\end{table*}%

\subsection{Discussion and Limitations}
\label{Discussion}
Our results endorse the use of LPs to pick the cut-off layer when compressing LLMs by means of layer pruning. Considering the size of the LLMs at stake, as well as the comprehensiveness of the three chosen datasets, we believe that the validity of our findings is supported. In fact, the results obtained by compressed LLMs outperform the state of the art, as it will be further shown in Section \ref{RW}. Interestingly, our compressed models improve the results as compared to the original, non-compressed versions. As pointed out in \cite{zhou2024survey}, model-level optimizations such as compression or pruning tend to come with a loss in performance. Nonetheless, our results show that pruning a model for a specific task and optimizing it for a really specific downstream task, such as code vulnerability detection, can even lead to better results.

In the same vein, LPs have been shown to be effective for having an early (yet accurate) estimate on the post-fine-tuning and post-compression model performance. Estimation errors are affordable, indeed. However, we opted for using a linear relationship between LP results and the target values (recall Equation \ref{estimationa}), as the amount of instances is limited. Therefore, exploring non-linear relationships could lead to better results and is an interesting research direction.

LPs are also affordable and cost-effective. The training time per layer for the LPs in BERT is, on average, 0.05 seconds per layer across the three datasets, compared to the 6,072 seconds it takes on average to fine-tune BERT across the datasets. For Gemma, only 1.49 seconds per layer is needed for LPs, in contrast with the 5,162 seconds required on average to fine-tune the model. For each sample, the time required to extract the features is 0.007 seconds for BERT and 0.009 seconds for Gemma. While the final time required to compute a single round of the LPs may depend on the number of samples used, 98.18, 21.92, and 41.69 seconds were required with BERT for DiverseVul, PrimeVul, and BigVul, respectively. For Gemma, 142.97, 46.71, and 79.04 seconds were needed for the same datasets. This represents a reduction of 99.11\% and 98.26\% in the average time needed to fine-tune the models, respectively.

On the contrary, our results are not enough to confirm that LPs can be directly applied to other domains (such as speech recognition or NLP-related tasks). Indeed, choosing which LPs and features to apply is not immediate. Hence, this work must be regarded as a first attempt in this direction, opening interesting research venues to ascertain their use in other domains or with different LLMs. %Indeed, the use of LPs to determine the most convenient layer for pruning has shown to be effective for vulnerability detection, but choosing which LPs to implement in other areas  is far from straightforward. 
  
Our experimental settings impose a number of natural limits to our results. Thus, only C/C++ programming languages are at stake, but additional ones could be tested. Similarly, we have resorted to a classic LLM (Bert) and a very novel one (Gemma). However, others like Gamma or GPT-4o, could also be relevant.

%{\color{red} 
%Luis: Possible biases of the use of GaLore, low rank, etc.
%}

%{\color{red} 
%Luis: Why not merging layers? or doing anything else, e.g., knowledge distillation? why didn't we choose them, which are the hard issues related to them, possible barriers of adoption, etc.

%}

%{
%\color{black}
%To compare whether performing quantization after fine-tuning affects the models, we used Galore, the only available methodology that allows near full fine-tuning in a reasonable time on a consumer-grade GPU without replacing the linear layers with the quantized version during training. For example, QLORA results in a model with already quantized layers, making it unsuitable for comparison with BERT.

Regarding other compression methodologies, such as merging layers or knowledge distillation, we consider these techniques complementary to our approach. We could use a larger teacher model to enhance the representations of our pruned models or further compress our pruned model by merging similar layers. These are interesting lines of research to explore if LPs can be used in addition to these methods to create smaller models while leveraging the insights from the hidden states of the LLMs.

%In terms of energy, the different set-ups forces the GPUs to go to maximum capacity, which is 420 Watts. To make a fair comparison between base and pruned models, same batch sizes should have been used.

%}
%Another issue is the types of used CWE, which correspond to the 10 CWE with more samples among the Top25 more dangerous ones. This is a fair assumption but more CWE could be included, though dealing with balance problems in different datasets.

%The selection of classes for MLPs in linear probes has been enforced according to the right predicted class. However, predicting a contiguous class is different than a separated one, e.g. if the class is 4, the prediction of 5 or 8 presents quite disparate meaning. This fact could be part of the analysis of classes' distribution and may affect their final specification. 

\section{Related work}
\label{RW}

\begin{table*}[htbp]
  \centering
  \caption{Vulnerability detection using LLMs. Notice column Binary (B)/ Multiclass (M) marks if the set-up was binary or multiclass.}
  \scalebox{0.55}{
    \begin{tabular}{|p{7.39em}|p{6em}|p{10.39em}|p{16.39em}|p{10.39em}|p{5.39em}|p{4.39em}|p{7.39em}|c|p{7.39em}|}
    \hline
    \textbf{Reference} & \textbf{Programming language} & \textbf{Dataset} & \textbf{Model} & \textbf{Results} & \textbf{CWE-level analysis} & \textbf{Binary (B)/ Multiclass (M)} & \textbf{Input features} & \multicolumn{1}{p{3.055em}|}{\textbf{Max tokens}} & \multicolumn{1}{p{3.055em}|}{\textbf{Compression technique(s)}} \\
    \hline
        Fu et al \cite{fu2022linevul} & C/C++ & Big-Vul & Custom transformer & \multicolumn{1}{c|}{ Big-Vul: 65\% Acc.} & $\checkmark$    & B     & Code  & 512 & $\times$ \\
    \hline
    Steenhoek et al \cite{steenhoek2023empirical} & C/C++ & Devign, MSR & Vulberta and other custom transformers models & Devign: 55\%-89\% F1 & $\times$    & B     & Code  & 512 & $\times$ \\
    \hline
    Ahmad et al. \cite{ahmad2023flag} & C/C++, Verlilog, Python & Custom dataset & Codex, GPT-3.5 & \multicolumn{1}{c|}{Custom dataset: 66\% Acc.} & $\checkmark$    & B     & Code + prompt & \multicolumn{1}{p{3.055em}|}{50 lines} & $\times$ \\
    \hline
    Zhou et al. \cite{zhou2024large} & C/C++ & Vulnerability-fixing commit dataset & GPT-3.5 GPT-4 & VFC: 75.5\% Acc. & $\times$     & B     & Prompt + code & 4,096 & $\times$ \\
    \hline
    Fu et al. \cite{fu2023chatgpt} & C/C++ & Big-Vul & Chat-GPT & Big-Vul with prompts: 13\%-20, M. 10\% Acc., B\%  
    
    Big-Vul fine-tuned LLM: 65\% Acc., M. 94\% Acc., B& $\times$     & B and M & Prompt + code & 4,096 & $\times$ \\
    \hline
    Fu et al. \cite{fu2023vulexplainer} & C/C++ & Big-Vul & CodeBERT, Devign, ReGVDm GraphCodeBERT, LFME, BAGS &Big-Vul: 64\% Acc. & $\checkmark$     & M     & Code  & 512 & $\times$ \\
    \hline
    Du et al. \cite{du2024vulnerability} & C/C++ & Big-Vul & GraphCodeBert &Big-Vul: 93.83 \% Acc. & $\times$     & B     & Code  & 512 & $\times$ \\
    \hline
    Chen et al. \cite{chen2023diversevul} & C/C++ & diversevul, Devign, ReVeal, Big-Vul, CrossVul, CVEFixes (Jointly) & Roberta, GPT-2, T5 & 
    CVEFixes: 91.64 \% Acc.

    Devign, ReVeal, Big-Vul, CrossVul: 92.30\% Acc.

    DiverseVul: 92.30 \% Acc. (Best effort) & $\checkmark$     & B     & Code  & 512 & $\times$ \\
    \hline
    Gao et al. \cite{gao2023far} & C/C++ & d2a, ctf, magma, big-vul, and devign & chatglm2 -6b Llama-2 -7b vicuna -7b vicuna -7b-16k Llama-2 -13b Baichuan2 -13b vicuna -13b vicuna -13b-16k internlm -20b vicuna -33b CodeLlama -34b falcon -40b Llama-2 -70b Platypus2 -70b GPT-3.5 GPT-4 & Joint datasets:  40.6 \% F1, B / 37.9 \% F1,M & $\checkmark$     & B and M & Prompt + code & 2,048 & $\times$ \\
    \hline
    Ding et al. \cite{ding2024vulnerability} & C/C++ & Primevul & Codet5, codebert, unixcoder, starcoder2,codegen, gpt-3.5 Gpt4 &PrimeVul: 96\% Acc. & $\times$     & B     & Code  & 512 & $\times$ \\
    \hline
    Hanif et al. \cite{hanif2022vulberta} & C/C++ & Pre-training (GitHub,Draper) + finetuning on Vuldeepecker, Reveal, Draper, muVuldeepecker, D2A,Devign & Vulberta &Joint datasets: 99.59\% F1 & $\checkmark$     & M     & Code  & 512 & $\times$ \\
    \hline
    Tamberg et al. \cite{tamberg2024harnessing} & Java  & Java Juliet 1.3 & CodeQL, GPT-4, CLaude &Java Juliet: 72\% (max) Acc. & $\checkmark$    & B     & Code  & \multicolumn{1}{p{3.055em}|}{-} & $\times$ \\
    \hline
    Shestov et al. \cite{shestov2024finetuning} & Java  & VCMatch (custom) & WizardCoder, coderbert & VCMatch: 75\% - 85\% ROC & $\times$    & B     & Code  & \multicolumn{1}{p{3.055em}|}{2,048  and  512} & $\times$ \\
    \hline
    Jensen et al. \cite{jensen2024software} & Python & HumanEval, MBPP, SecurityEval 660 between all & Falcon-7b, Llama, llama2, dolly & Joint datasets: 95.6\% Acc. with 37.9\% F1 & $\times$    & B     & Prompt + code & 4,096 & $\times$ \\
    \hline
    Shi et al. \cite{shi2022compressing} & C/C++ & Devign & CodeBERT &Devign: 59\% Acc.  & $\times$     & B     & Code & 512 & $\checkmark$ Knowledge distillation \\
    \hline
    \textbf{Ours} & C/C++ & Big-Vul, DiverseVul, PrimeVul & BERT, Gemma & Big-Vul: 82\%-96\% Acc.
DiverseVul: 77.7\%-96\% Acc.
PrimeVul: 87.1\%-99\%  Acc. & $\checkmark$    & B and M & Code & 1,024 & $\checkmark$  Layer pruning and quantization, guided by Linear Probes \\
    \hline
    \end{tabular}%
    }
  \label{tab:RW_vul}%
\end{table*}%

Vulnerability detection using LLMs presents a critical challenge in the development of tools that assist security analysts in writing or auditing code. The vast array of existing vulnerabilities, including zero-day threats, exacerbates this challenge due to the potentially infinite spectrum of vulnerabilities \cite{zhout2023devil}.

For this reason, academia has extensively explored various proposals. General models such as GPT-4, Chat-GPT, and LLama2 have been studied \cite{fu2023chatgpt,zhou2024large} to evaluate their effectiveness in classifying vulnerabilities across binary and multi-class frameworks \cite{gao2023far}, specifically within C, Java \cite{tamberg2024harnessing}, and Python code \cite{jensen2024software}, utilizing diverse prompting strategies. Ahmad et al. \cite{ahmad2023flag} studied these strategies to find vulnerabilities at a line-level approach, showing promising results.

However, despite these efforts, the studies indicate that the current state of the art in prompt-based models underscores the necessity for specialized models tailored for vulnerability detection, particularly as current models struggle with accurately identifying the potential CWE of the code (see \cite{yao2024survey} for more).

Among specialised models, finetuned LLMs have been the option that has lead to better results. Chen et al. \cite{chen2023diversevul} covered a wide range of models' families (RoBERTa,T5,GPT-2) while introducing a new diverse C code vulnerability dataset. Fu et al. \cite{fu2022linevul} studied the possibility of predicting vulnerabilities on a line based approach. However, the problem was covered as a binary classification problem and thus, models lack the ability of correctly discerning the CWE type. They concluded that diversity of data sources, vulnerabilities and pre-training on code data are crucial for better detection while larger models seems not to be beneficial. This was later also pointed by Steenhoek et al. \cite{steenhoek2023empirical} who carried out an empirical study for vulnerability detection with fine-tuned LLMs.

Hanif et al. \cite{hanif2022vulberta} explored to pre-train and finetune a model (RoBERTa) specifically for code vulnerability detection. They reach good results in both, binary and multi-class vulnerability detection. Nonetheless, pre-training this model requires 96 hours and 80 GB of VRAM, which are usually resources not generally accessible. Additionally, no vulnerable samples are ignored in the milticlass approach, training the model for just distinguishing among the different CWEs. %the class in multi-class for non-vulnerable is ignored, training model just on distinguishing among the different CWEs. 
The characterization of CWE was covered by \cite{fu2023vulexplainer} who proposed a set of abstract classes to reduce the complexity of fine-grained predictions by grouping similar CWE.

Among compression proposals, only Shi et al. \cite{shi2022compressing,shi2024greening} have addressed the topic. They propose a two-step method to find optimal model configurations for a base size BERT model (CodeBERT). First, they maximize accuracy while reducing model size using a genetic algorithm. Afterward, they apply knowledge distillation to the optimized model. However, this approach results in an accuracy of only 59\% for binary code vulnerability detection, which is close to random guessing. The trial-and-error methodology for testing configurations could be impractical due to the complex hyperparameter search space, especially for larger models like those discussed in our paper.

Their solution requires a notable amount of time. In the first step, they need to limit the search space, which takes 5 minutes on a cluster with 80 CPUs and 504 GB of RAM. The resultant set of 20 models needs to be pre-trained, taking a minimum of another 10 hours. Finally, the fine-tuning process must be carried out, requiring an additional 20 minutes. This process is both time and energy-consuming. As pointed out in \cite{wang2023energy}, while fine-tuning consumes significant energy and generates emissions, pre-training the models is the most resource-intensive step. In our solution, no additional pre-training is required, thereby saving both energy and time.

Interestingly, our results suggests that compression techniques can also lead to improved performance results, as opposed to the findings by Shi et al. which characterize the incurred loss. 
%Our approach encompasses a broad array of datasets and is capable of not only distinguishing among various CWEs but also determining whether code snippets are not vulnerable. Moreover, the proposals for fine-tuned models are limited to 512 tokens, which is the maximum context window supported by those models. Additionally, we have explored the internal representation of different CWEs using linear probes to subsequently prune and quantify the model, an idea previously unexplored in this field.

Table \ref{tab:RW_vul} presents a comparison among works leveraging LLMs for vulnerability detection, pointing out the language of the code, datasets, models, main results, if the study is carried out in a binary or multi-class classification considering CWE or not, together with input features and tokens size. $LPASS$ encompasses three C/C++ datasets, the most common programming languages, and distinguishes among various CWEs, doing a multi-class classification and determining also whether code samples are not vulnerable. The average accuracy that we achieve across datasets for multi-class classification is 81\% for BERT and 76.6\% for Gemma, both of which outperform other approaches reported in the literature \cite{fu2023chatgpt,fu2023vulexplainer}. Specifically, on the Big-Vul dataset, BERT and Gemma achieve accuracies of 64\% and 65\%, respectively, which are lower than our best result of 82\%.
For binary classification, our results can be compared with studies that employ related datasets \cite{chen2023diversevul,du2024vulnerability,fu2022linevul,ding2024vulnerability,fu2023chatgpt}. On the Big-Vul dataset, our approach achieves an accuracy of 96\%, surpassing the results of \cite{du2024vulnerability,fu2022linevul,fu2023chatgpt}, which reported accuracies ranging from 65\% to 94\%, with a maximum of 93.83\%. Additionally, the work by Chen et al. \cite{chen2023diversevul} on DiverseVul yields an accuracy of 92.30\%, which is lower than our performance of 96\%. Finally, \cite{ding2024vulnerability} reports an accuracy of 96\% on the PrimeVul dataset, while our approach achieves 99\%. However, proposals in \cite{tamberg2024harnessing,shestov2024finetuning,jensen2024software} are not comparable to our work, as they focused on Python and Java. Additionally, while other efforts such as \cite{shi2022compressing,hanif2022vulberta,gao2023far,zhout2023devil,ahmad2023flag} did focus on vulnerability detection in C, like our study, they use datasets like Devign or others that only indicate whether a vulnerability is present or not and then, as pointed out in Section \ref{Datasets}, their use has been discarded in our work.

Moreover, proposals for fine-tuned models  \cite{hanif2022vulberta,chen2023diversevul,fu2023vulexplainer} are limited to 512 tokens, which is the maximum context window supported by most models.By contrast, we test a well-known model, BERT, and Gemma, not used until now, with 1,024 tokens in this latter case and using code as input features. 
Additionally, the model at stake in the work by Shi et al.  \cite{shi2022compressing} is already smaller than ours. Thus, it is unknown how well their proposal may work with larger models such ours as their approach may result in a substantially larger search space for their algorithms.

\section{Conclusion}
\label{sec:conclusion}
Reducing the size of LLMs is critical to ensure their scalability for vulnerability detection -- as real-time demands come into play, saving time and memory is paramount. However, model compression and fine-tuning requires non-negligible resources, and the effect of these actions into the model performance is unknown beforehand. To address these issues, in this paper an approach (dubbed $LPASS$) has been proposed to assist in making informed decisions. $LPASS$ helps on selecting the cut-off point for model compression using layer pruning. Moreover, it provides a good estimate of the post-fine-tuning and post-compression performance by using linear classifier probes. Indeed, $LPASS$-based versions of two LLMs have not only outperformed the state-of-the-art in vulnerability detection, but also non-compressed versions, thus showing the suitability of using probes.   %different techniques of model compression have been used. Remarkably, the use of linear classifier probes has shown to be effective to provide an early estimate of the LLM results after fine-tuning and compression. All in all, the compressed versions of well-known LLMs has shown to be largely effective for the most dangerous vulnerabilities. 

Our results open a number of future research directions. On the one hand, our use of linear classifier probes had not been proven in the context of model compression. Thus, the analysis on their suitability for other models or application domains is relevant. On the other hand, in the field of vulnerability detection, the suitability for other languages remains an open issue.

\section*{Declarations}

\subsection*{Ethical Approval and Consent to participate}
\noindent The authors declare they have no conflict of interest.

\subsection*{Consent for publication}
\noindent Not applicable

\subsection*{Availability of supporting data}
\noindent Datasets are public. Our experimental code will be freely available if the paper is accepted for publication.

\subsection*{Competing interests/Authors' contributions}
\noindent The authors declare that they have no competing interests.

\subsection*{Funding}
%Hidden for anonymity purposes
Nicolas Anciaux was supported by the French grant \href{https://www.pepr-cybersecurite.fr/projet/ipop/}{iPoP} PEPR (ANR-22-PECY-0002). Luis Ibanez-Lissen was supported, and also Lorena Gonzalez partially, by the Spanish National Cybersecurity Institute (INCIBE) grant APAMciber within the framework of the Recovery, Transformation and Resilience Plan funds, financed by the European Union (Next Generation). Jose Maria de Fuentes was partially supported by grant PID2023-150310OB-I00 of the Spanish AEI. Jose Maria de Fuentes and Lorena Gonzalez have also received support from UC3M's Requalification programme, funded by the Spanish Ministerio de Ciencia, Innovacion y Universidades with EU recovery funds (Convocatoria de la Universidad Carlos III de Madrid de Ayudas para la recualificación del sistema universitario español para 2021-2023, de 1 de julio de 2021).

\bibliographystyle{elsarticle-num}
\bibliography{bibFile}

\end{document}